\documentclass[referee,traditabstract]{aa}

\usepackage{graphicx}
\usepackage{epsfig}

\usepackage{txfonts}

\usepackage{natbib}
\bibpunct{(}{)}{;}{a}{}{,}

\begin{document}

\title{The Evolution of the EM Distribution in the Core of an Active Region}

\author{Giulio Del Zanna \inst{1} 
		\and Durgesh Tripathi \inst{2} 
\and Helen Mason \inst{1}
		\and Srividya Subramanian \inst{2} 
		\and Brendan O'Dwyer \inst{1} 
		}

\institute{Department of Applied Mathematics and Theoretical Physics, University of Cambridge, Wilberforce Road, 
			Cambridge CB3 0WA, UK \label{inst1} 
			\and 
			Inter-University Centre for Astronomy and Astrophysics, Post Bag-4, Ganeshkhind, Pune 411007, India \label{inst2}
			} 

\date{Received <date> / Accepted <date>}

\abstract{We study the spatial distribution and evolution of the 
slope of the Emission Measure
between 1 and 3~MK in the core  active region NOAA~11193, 
 first when it appeared near the central meridian and then again when it 
re-appeared after a solar rotation. 
We use  observations recorded by the Extreme-ultraviolet Imaging Spectrometer (EIS) aboard Hinode, with a new 
radiometric calibration. We also use  observations from the Atmospheric Imaging Assembly (AIA) 
aboard Solar Dynamics Observatory (SDO).
We present the first spatially resolved maps of the
EM slope in the 1--3~MK range 
within the core of the AR using several methods, both
approximate and from the Differential Emission Measure (DEM).
A significant variation of the slope is found at different spatial locations
within the active region.
We selected two regions that were not affected too much by
any line-of-sight lower temperature emission. We  found that the EM had a power law
of the form EM~$\propto T^{b}$, with b = 4.4$\pm0.4$, and 4.6$\pm0.4$,
during the first and second appearance of the active region,
 respectively.  During the second rotation, 
line-of-sight effects become more important, although  difficult to estimate. 
We found that the use of the ground calibration for Hinode/EIS and the approximate method to derive the Emission
Measure, used in previous publications, produce an underestimation of the
slopes.  The EM distribution in active region 
cores is generally found to be consistent with high frequency heating,
and stays more or less the same during the evolution of the active region. 
}
\keywords{Sun: atmosphere -- Sun: corona, Methods:observational -- Methods: data analysis}
\titlerunning{EM distribution as function of AR age}
\authorrunning{Del Zanna et. al.}

\maketitle
\section{Introduction}

The solution to the long standing problem of solar coronal heating remains elusive despite major advances in observational and theoretical 
capabilities over the last few decades (see \citealp{Kli:06} for a review). 
We now know that active regions generally comprise a variety 
of structures which  are broadly classified as warm loops [$T\simeq$ 1
MK], fan loops [$T<$~0.8 MK],  
mostly emanating from  sunspots, and hot loops [$T\simeq$ 3 MK] in the cores of active regions. Additionally, 
it is also known that there is significant unresolved emission in the 1--3 MK range 
 (see e.g. \citealp{DelM:03}, \citealp{ViaK:12},  \citealp{delzanna:2013}, \citealp{SubT:14}). 
Any theory for coronal heating must explain the emission from these different kinds of loops 
as well as the diffuse emission.

A clear understanding of the thermal distribution and time scale of energy release in coronal structures reveals information regarding the heating 
mechanisms in that particular structure. There is an on-going debate in the current literature as to whether the heating is low- or high-frequency. 
High (low) frequency heating occurs when the duration between successive heating events is smaller (larger) than the cooling time. 
In the high-frequency heating scenario, the plasma does not have enough time to cool sufficiently and 
produces  a narrow emission measure (EM) distribution with a 
steep slope  $b$  [EM(T)~$\propto$~T$^b$]
in the 1--3~MK range. However, in the low frequency heating scenario, 
since the time duration is larger than the cooling time, the plasma 
has sufficient time to cool down before being re-heated again. 
Hence, there would be substantial amount of material at cooler temperatures giving 
rise to comparatively shallower slopes (see, e.g. 
\citealp{mulu-moore_etal:2011,TriKM:11,bradshaw_etal:2012,cargill:2014} 
and references therein).
Generally, the models predict  that
low-frequency nanoflares can only account for slopes $b$  that are below 3.

The  emission of the  $\simeq$ 1~MK  `warm' loops observed in EUV with the Extreme-ultraviolet Imaging Telescope 
(EIT; \citealp{DelEtal:95}) on board SoHO, the Transition Region And Coronal Explorer (TRACE; \citealp{HanEtal:99}), the Extreme-Ultraviolet 
Imaging Spectrometer (EIS; \citealp{CulEtal:07}) abroad Hinode and most recently by the Atmospheric Imaging Assembly (AIA; \citealp{LemEtal:12}) 
on board the Solar Dynamics Observatory (SDO) is generally explained by low frequency heating
(see e.g. \citealp{WarWM:03},\citealp{WinWS:03}, \citealp{WarU:08}, \citealp{TriMD:09}, \citealp{UgaWB:09}, \citealp{Kli:09}). 
The heating of the hot loops in the core of the active regions has been, however, a matter of much debate
(see e.g., \citealp{TriMK:10}, \citealp{WarWB:10}, \citealp{TriMD:10}, \citealp{TriKM:11}, \citealp{WarBW:11}, \citealp{WinSW:11}, 
\citealp{ViaK:11}, \citealp{TriMK:12}, \citealp{DadT:12}, \citealp{WarWB:12}, \citealp{SchP:12}, \citealp{UgaW:12}, \citealp{WinTMD:2013}).
One issue that is still not clear is the amount of hot plasma above 3~MK. 
EUV and X-ray spectroscopic observations indicate that the cores of quiescent ARs have very little 
hot plasma (see \citealt{delzanna:2013, delzanna_mason:2014} and references therein),
an issue that we do not discuss here.

This paper instead focuses on the slopes in the 1--3 MK temperature range.
Recent studies used  Hinode EIS 
 observations of the  cores of active regions, avoiding moss emission.
 For example, \citet{WarBW:11} studied an inter-moss region and found an EM 
distribution that can be approximated by EM~$\propto$~T$^{3.26}$.  
 For a different active region \citet{WinSW:11} found a similar power-law slope (EM~$\propto$~T$^{3.2}$).
 However, \citet{TriKM:11} found EM distributions with a power-law slope of approximately 2.4 
for several inter-moss regions of two active regions. A different method was used.   
\citet{WarWB:12}  studied the EM distribution in the cores of 15 active 
regions and found EM distributions that had a range of 
slopes, between 2 and 5. 

One question naturally arises: are these slopes different depending on the AR, and do they change during the 
lifetime of an active region?
The general evolution of an active region has been known for a long time from e.g.  
Skylab observations  (see  \citealp{sheeley:1981}),
however a quantitative analysis on small spatial scales was not available.
Recently, \citet{UgaW:12} studied two active region cores at several instances during their life time, 
with Hinode/EIS, STEREO/EUVI, and SDO/AIA. 
They found that the EM at 4 MK normally declines with time, following its suggested relationship with the magnetic flux 
\citep{WarWB:12}. However, they also found an enhancement with time in the EM at lower temperatures, 0.6-0.9~MK. 
Their results suggest that both low and high frequency heating can occur in active region cores 
during  the lifetime of an active region. 
The enhancement of EM at lower temperature could also be interpreted as dominance of low 
frequency heating events during the later part of an active regions' life. 
Similarly, \citet{SchP:12}  studied eight inter-moss regions in five different active 
regions (some overlapping with those studied by  \citealp{WarWB:12}). 
They combined Hinode EIS and XRT observations to better constrain 
the high temperature component of the EM. 
In addition they estimated the age of the active region, although they did not actually track the
same active region on two rotations. 
They concluded that their results were consistent with older active regions being more likely dominated
by steady heating (high frequency impulsive heating) and younger regions
 showing more evidence of low frequency impulsive heating.

We believe that it is important to track the evolution of the same active 
region.
In this paper we present  Hinode/EIS  and SDO/AIA observations of the core of the same active region 
at two instances during its life time, once when it appeared on 
disk and again when the active region was seen after one solar rotation.

Three questions naturally arise. 1) How do the slopes vary within the
active region? We discuss possible methods of obtaining the slopes for 
each observed pixel, from simple fast methods to more complex methods,
and we present the results.
2) How does the spatial resolution affect the results? 
We complement the  Hinode/EIS observations with the much
higher-resolution SDO/AIA data, and compare the slopes obtained 
from the two instruments. 
3) Given that different methods have been used in the literature to
derive the DEM and EM, how do the slopes depend on the method used?  
We compare the results obtained for different methods.

 \citet{guennou_etal:2013} has recently highlighted that atomic data 
uncertainties affect the 
estimation of the slopes. We show that another important issue 
that was neglected in previous literature is the EIS radiometric
calibration, which has
recently been revised  \citep{delzanna:13_eis_calib}.

The  paper is organised as follows:  in section~2 we describe the observations of the active region, 
in section~3 we review the different methods we adopt, and  in section~4 we 
describe our data analysis and results, followed by a summary and conclusions in section~5.

\section{Observations} \label{obs}

\begin{figure}[!htbp]
\centerline{
\epsfig{file=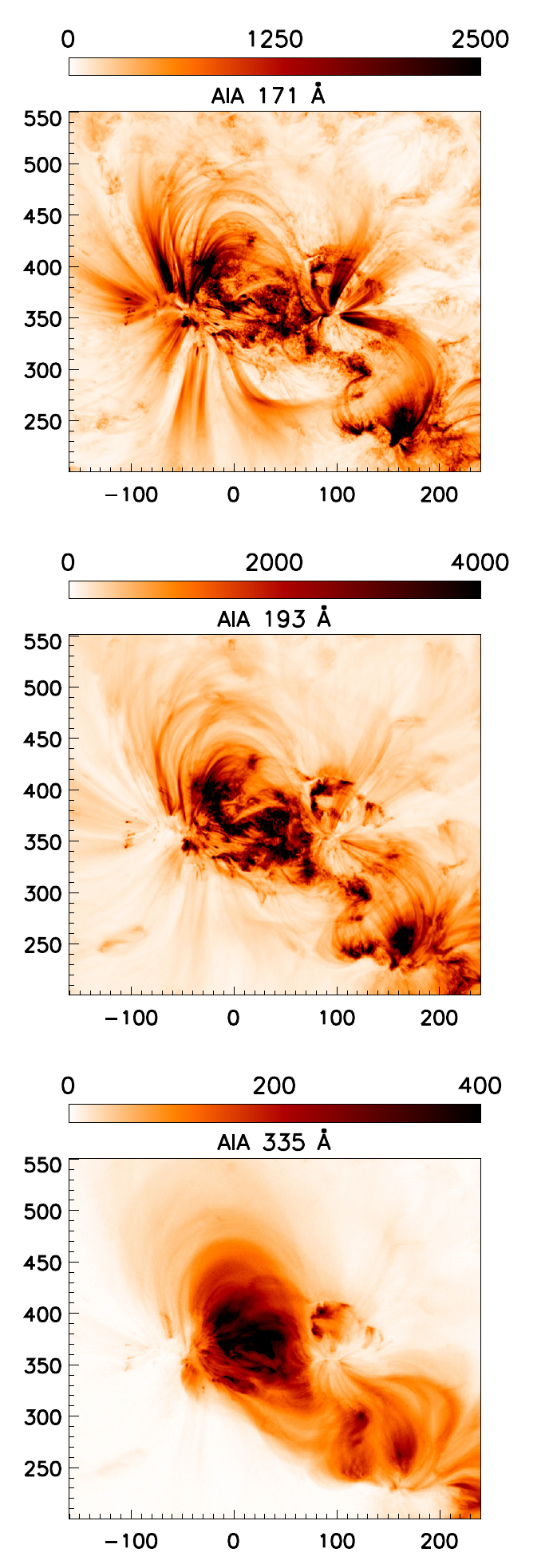, width=4.8cm,angle=0 }
\epsfig{file=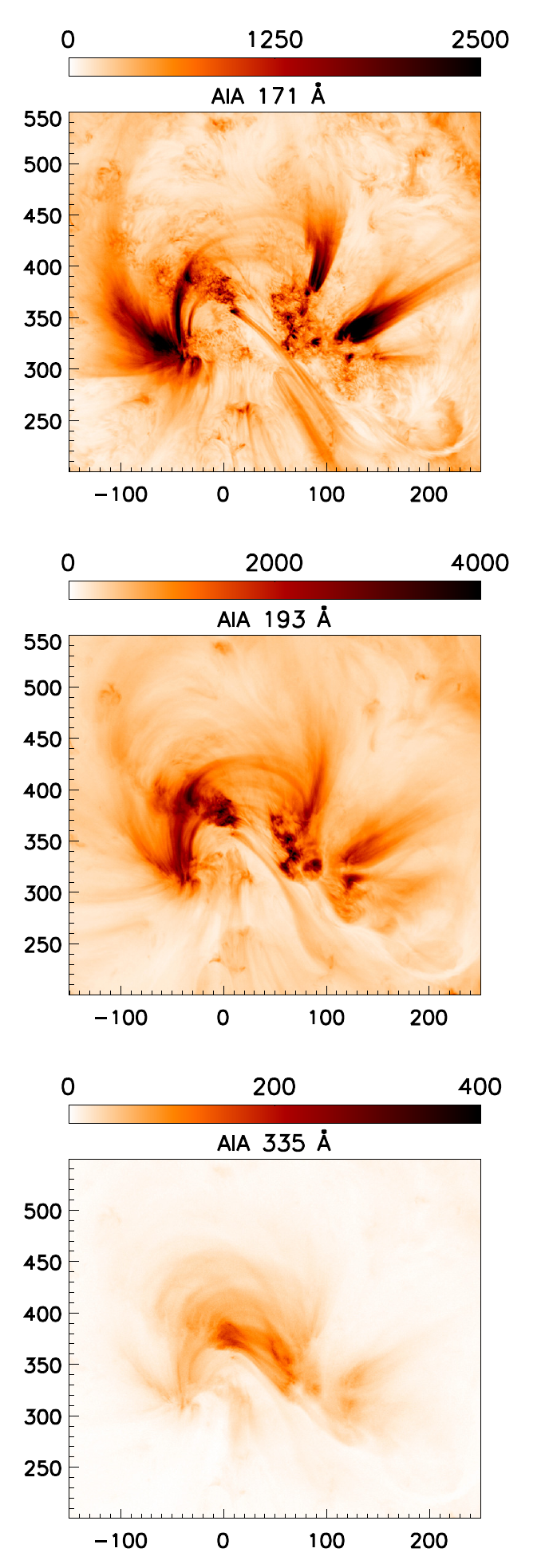, width=4.8cm,angle=0 }
}
\caption{AIA negative images of active region NOAA 11193  when it first appeared 
on the visible solar disk near the central meridian on Apr 19, 2011 (left) and 
during the second rotation on May 17, 2011 (right). The intensities are DN/s, and 
the images of the two dates are shown on the same intensity scale and field of view.
Coordinates are arcseconds from Sun centre. }
\label{aia_1}
\end{figure}

In the current analysis we have used observations recorded by EIS aboard Hinode and AIA aboard SDO. 
EIS provides high resolution 
spectra of the Sun in two wavelength bands, 170~-~211~{\AA},
 and 246~-~292~{\AA} \citep{CulEtal:07}, with an effective spatial resolution of about 3\arcsec.
AIA provides high-resolution imaging observations (with about 1\arcsec\ resolution,
0.6~arcsec/pixel) in 6 extreme-ultraviolet channels with high cadence \citep{LemEtal:12}. 

We  believe that line-of-sight effects can play an important role in estimating the slope of the EM, 
so we have searched for 
an active region that was observed  when it was close to the central
meridian, and were careful to  select  areas in the hot core loops which are free from contamination by 
low-lying moss regions. 
We have also analysed simultaneous SDO/AIA and Hinode/EIS observations to determine the EIS pointing and 
obtain high-resolution information.
For this study we have selected the active region NOAA~11193. This AR was first 
observed near the central meridian on Apr 19, 2011 and then again on May 16 and 17, 2011 
during its second passage from the central meridian on the visible solar disk. 
STEREO EUVI images indicate that the AR emerged on Apr 10, i.e. it was already 9 days old
at first meridian passage.

\begin{figure}[!htbp]
\centerline{\includegraphics[width=0.45\textwidth]{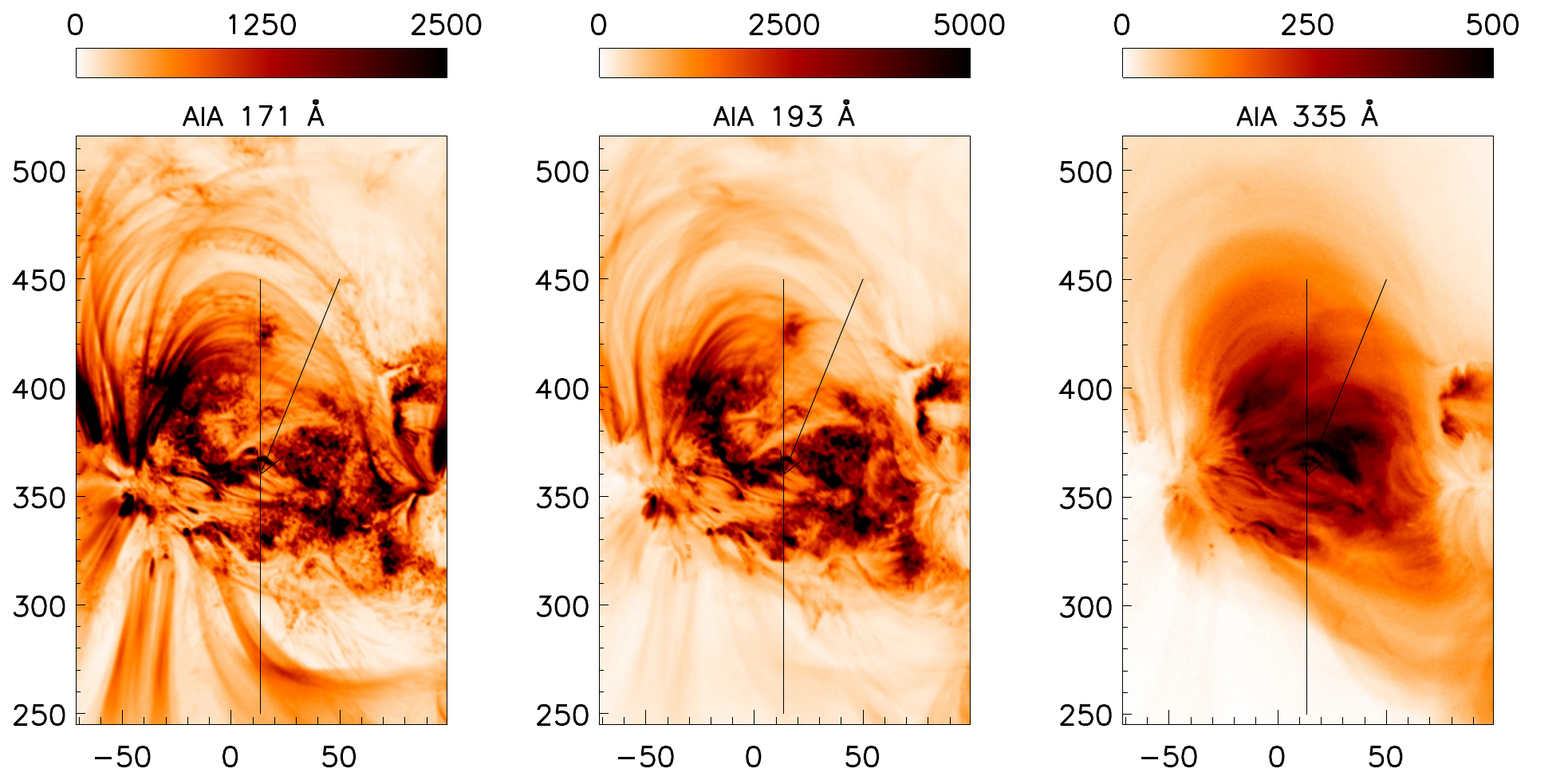}}
\centerline{\includegraphics[width=0.45\textwidth]{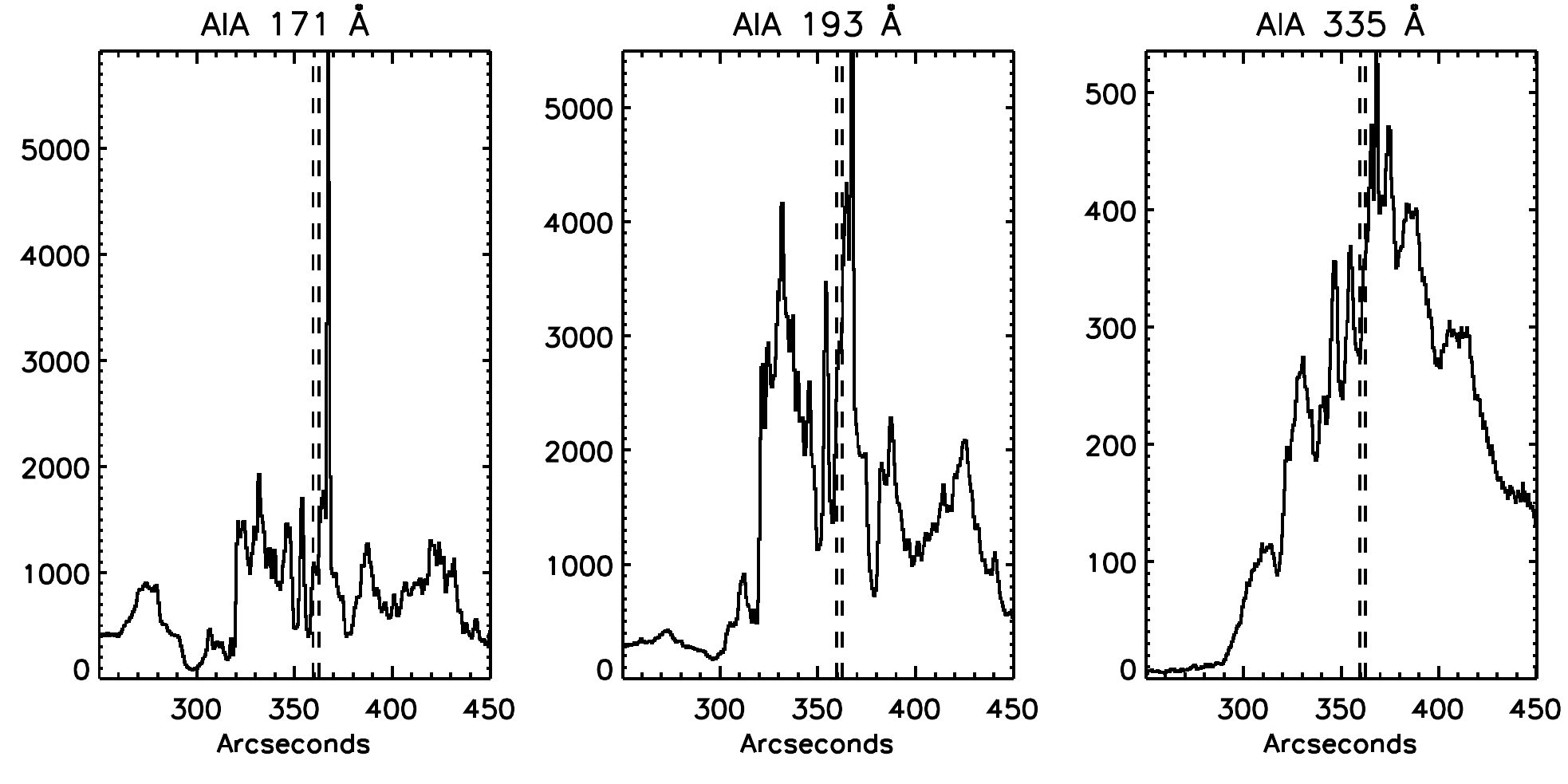}}
\centerline{\includegraphics[width=0.45\textwidth]{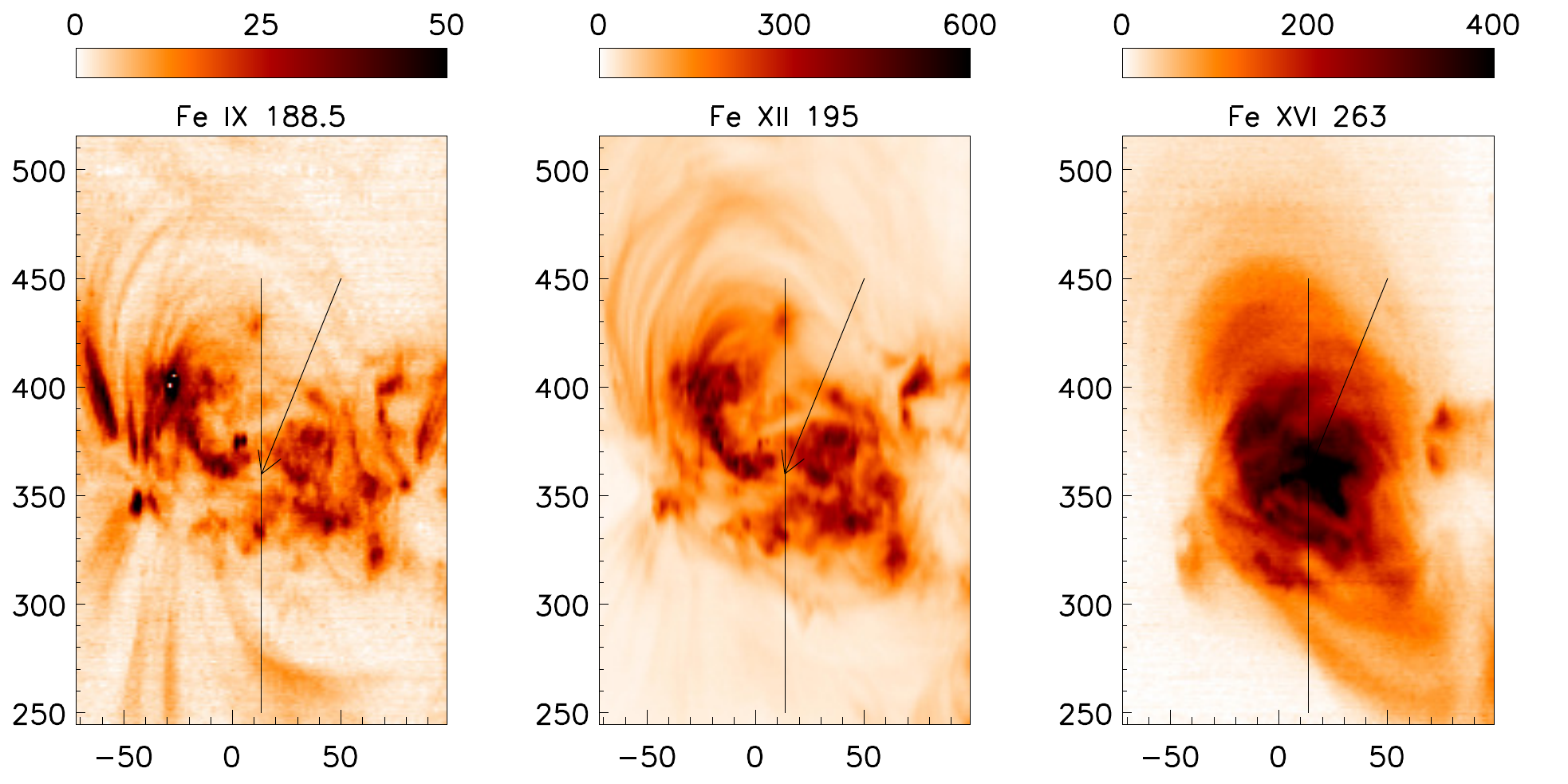}}
\centerline{\includegraphics[width=0.45\textwidth]{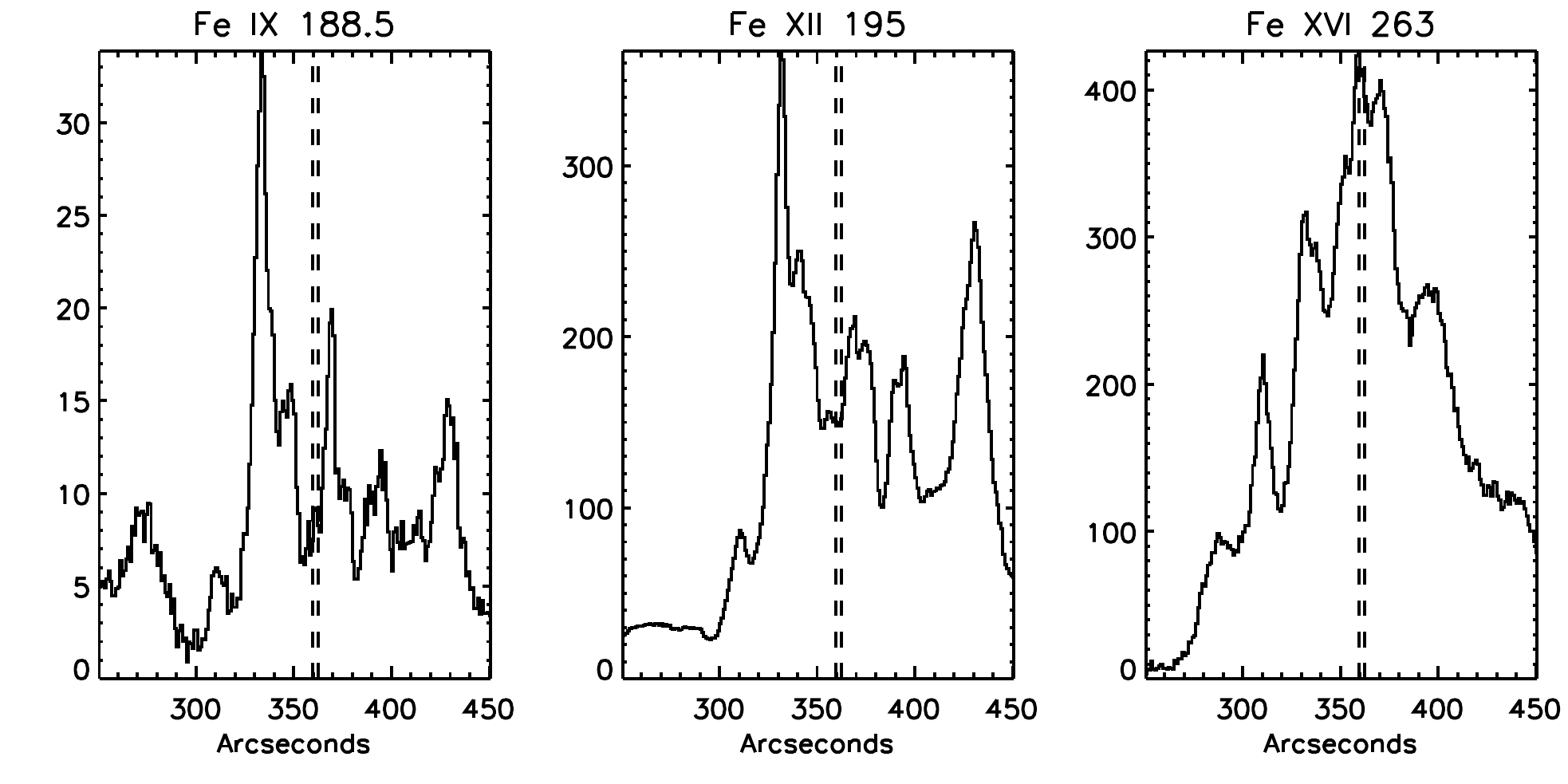}}
\caption{AIA (DN/s, first and second row) and EIS (radiances in phot cm-2 st-1 s-1,
third and fourth row) negative images  on Apr 19, 2011.
The arrows indicate the AR core region selected for DEM analysis.
The bottom plots show the profiles of the intensities along this vertical line.
The dashed vertical lines in the bottom plots 
indicate the  region selected for DEM analysis.
}
\label{aia_eis_19_apr}
\end{figure}

\begin{figure}[!htbp]
\centerline{\includegraphics[width=0.5\textwidth]{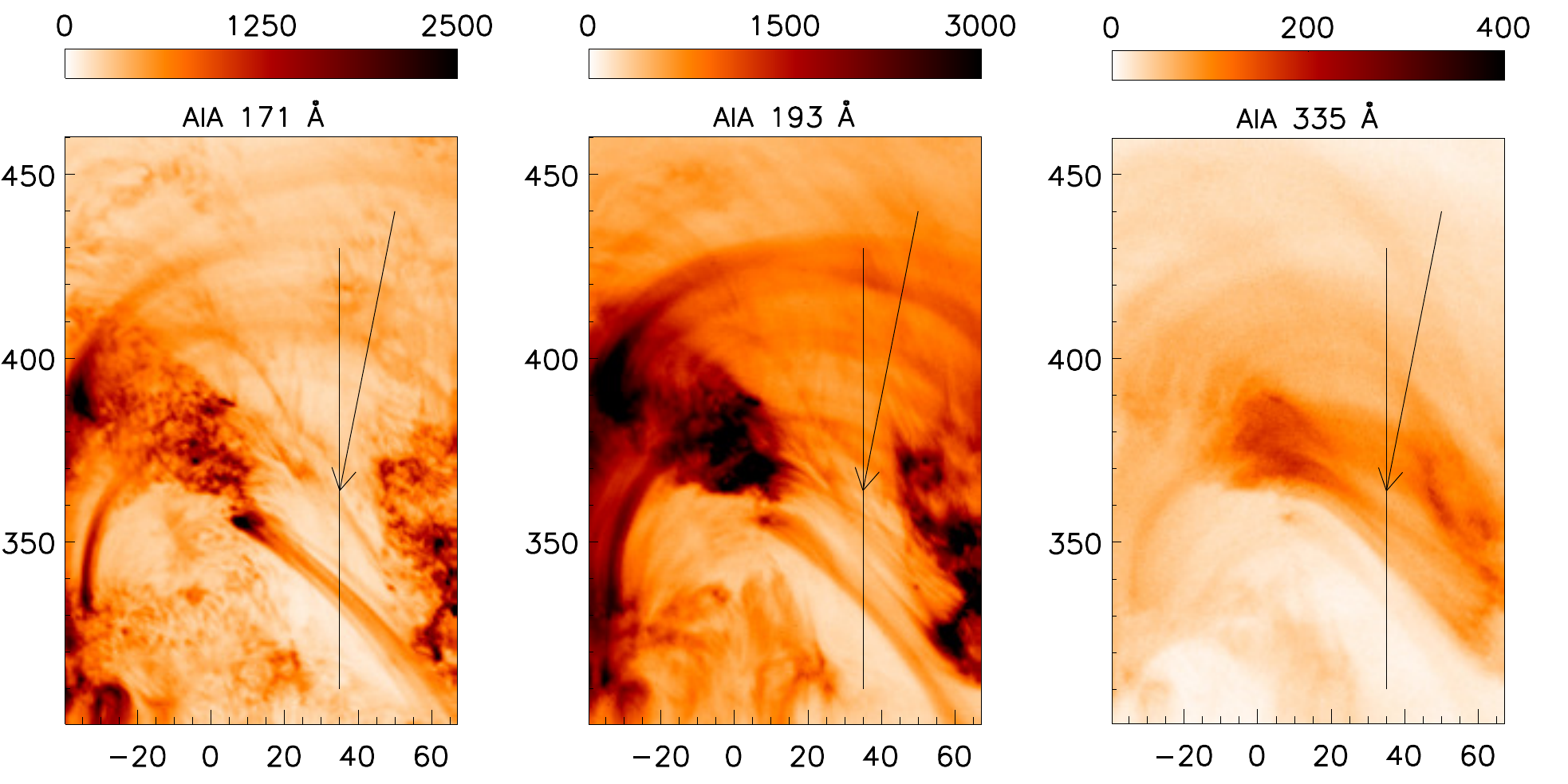}}
\centerline{\includegraphics[width=0.5\textwidth]{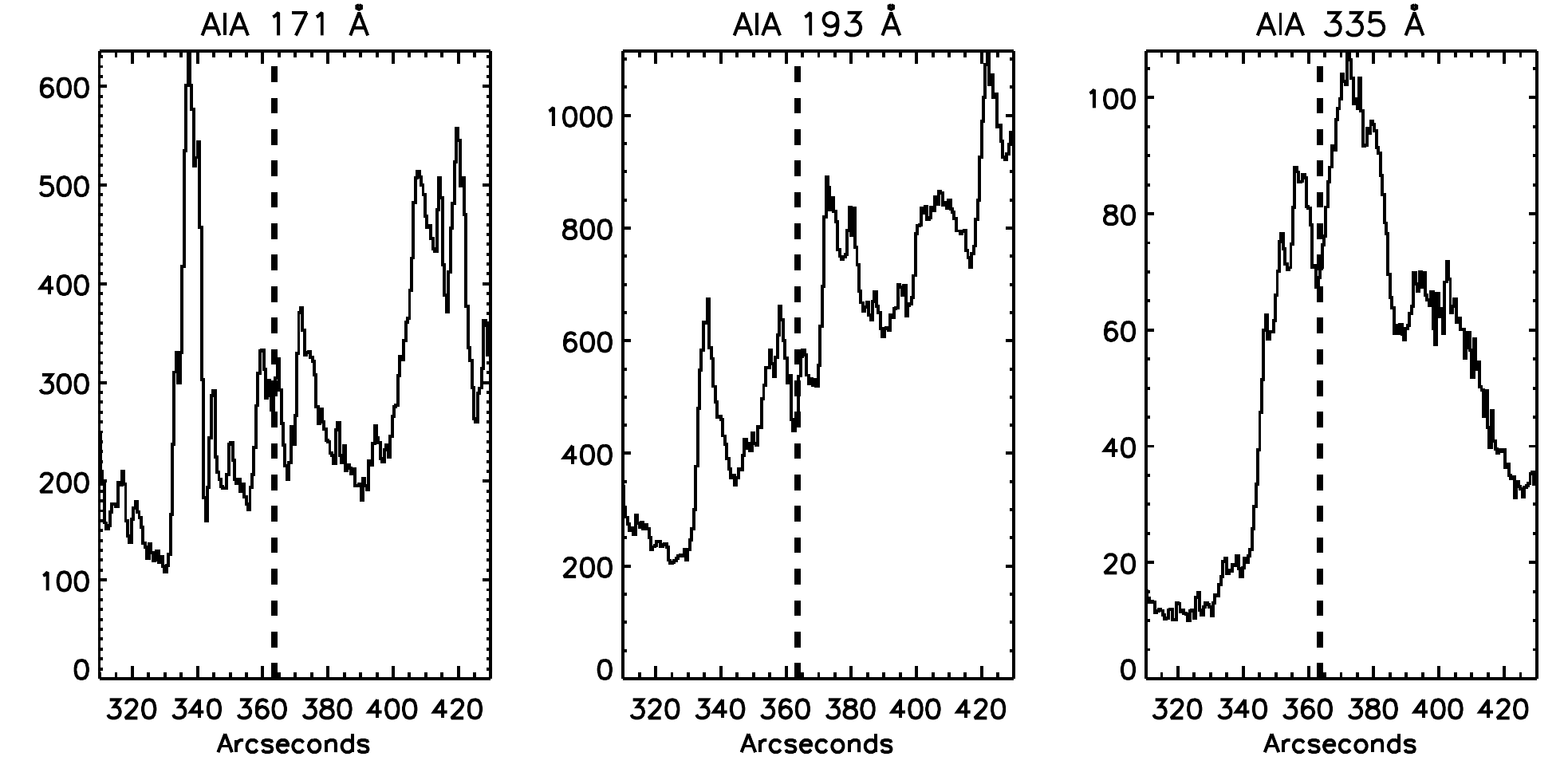}}
\centerline{\includegraphics[width=0.5\textwidth]{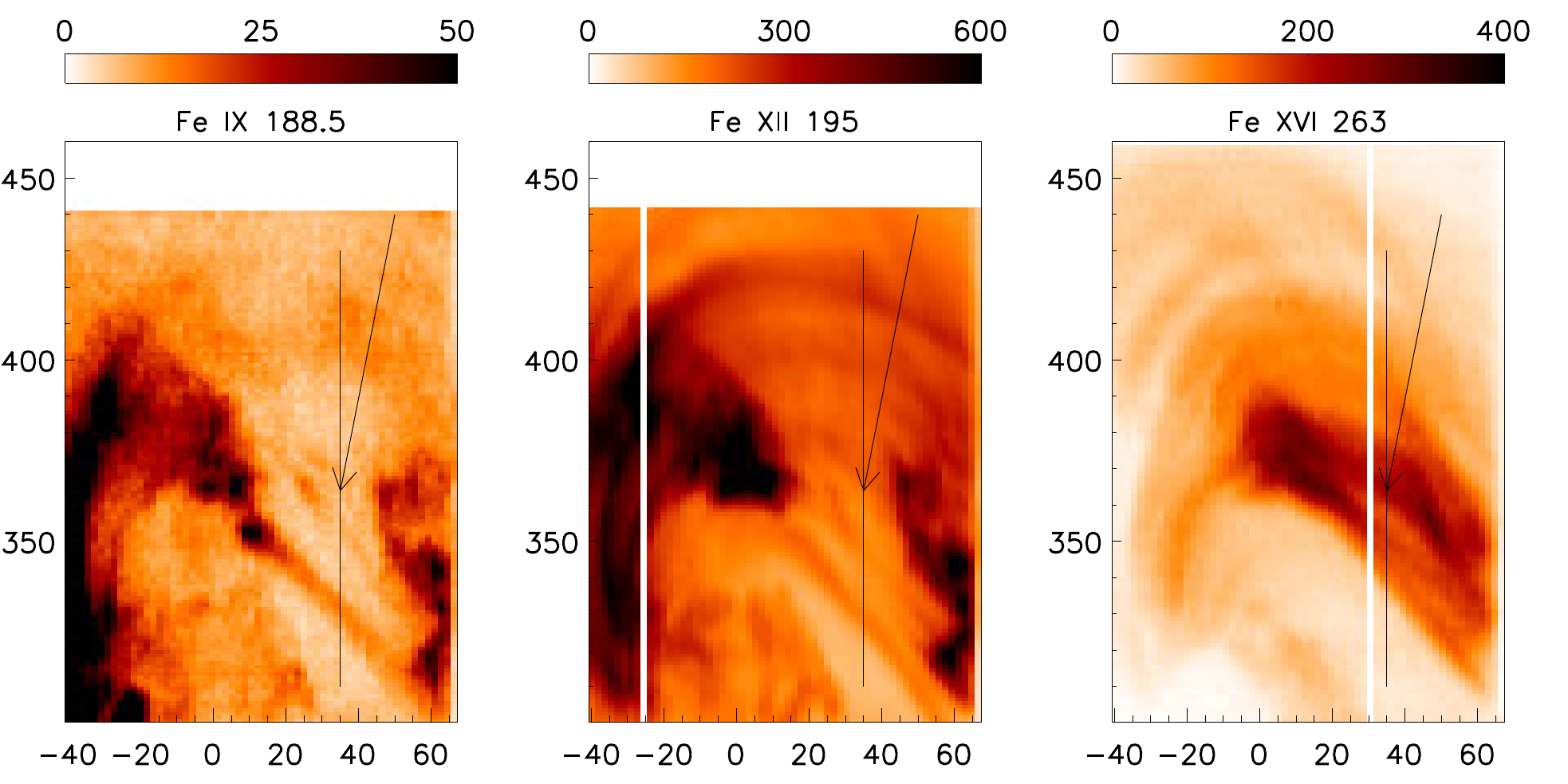}}
\centerline{\includegraphics[width=0.5\textwidth]{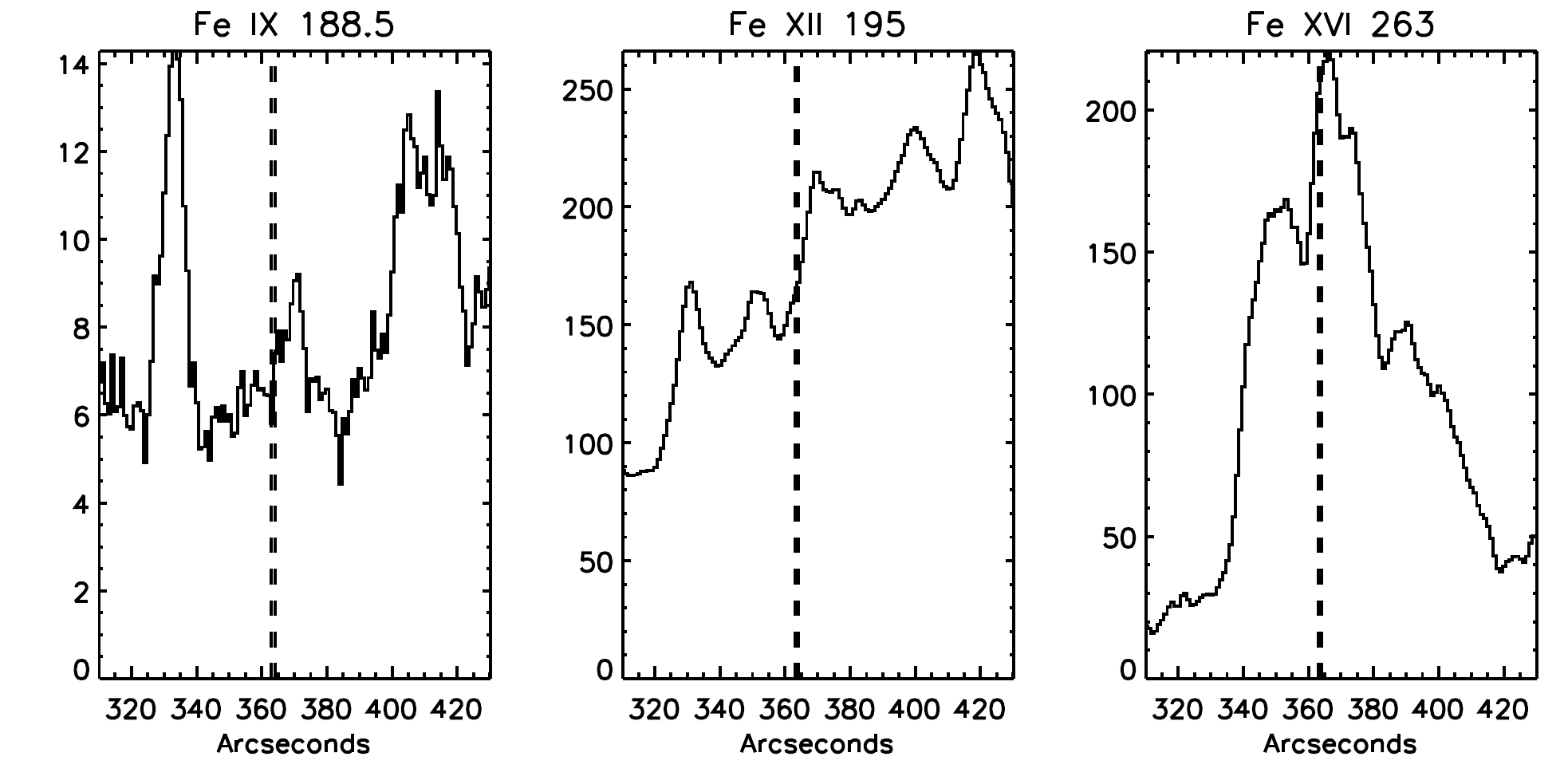}}
\caption{Same as Fig.~\ref{aia_eis_19_apr}, for the 17 May 2011 second rotation.}
\label{aia_eis_17_may}
\end{figure}

The AIA data have been processed in the following way
(see \citealp{DelOM:11} and \citealp{delzanna:2013}  for details).
We took the full-disk AIA data and adjusted the plate scale.
 We then corrected for the AIA stray light 
using the results of \citet{poduval_etal:2013}.
We note that the correction for the stray light  enhances the contrast 
between the loop structures and the background.
In order to align AIA and EIS accurately, the  AIA images were then 
reduced to the lower spatial and temporal resolution of the EIS data
for a direct comparison, i.e. to find the EIS pointing. 
The AIA and EIS comparisons
show, as in previous cases we have analysed  \citep{DelOM:11,delzanna:2013}, that the 
displacement in the E-W direction of the EIS slit between different exposures
is not equal to the nominal value.
For example, a step of 2\arcsec\ is in fact a step of 1.8\arcsec.

AIA images of the active region NOAA~11193 are shown in Fig.~\ref{aia_1}, 
when it was close to the central meridian on 
19 April 2011 and on 16 May 2011. The images are shown on 
the same intensity  scale to show how much the active region has changed.
The changes are typical of most active regions.
To actually interpret the emission in the SDO AIA bands, 
it is necessary to keep in mind that all the 
AIA bands are highly multi-thermal, as described in detail in
 \cite{OdwD:10,DelOM:11,delzanna:2013}.
The core of the AR in the 335~\AA\ band is dominated 
by the \ion{Fe}{xvi}~335.41~{\AA} line \citep{delzanna:2013}, which is formed at 3~MK, 
hence clearly shows the hot core loops.
There is a significant reduction in this 3~MK emission during the 
second rotation.
The brightest emission at 1--2 MK is in the low-lying moss regions,
the footpoint regions of most AR loops. The moss is clearly visible in the 
193~\AA\ band, which in these locations is dominated by \ion{Fe}{xii} emission 
\citep{delzanna:2013}.
The moss emission is significantly reduced during the second rotation.

The $\simeq$ 1~MK warm loops are already fully developed by the 19 April 2011, and 
are clearly visible in the 
 171~{\AA} channel, dominated by \ion{Fe}{ix}~171~{\AA}, which is 
formed over a relatively broad temperature range centered
around $\log\, T = 5.85$, as discussed in \citet{DelOM:11}.
The warm loops that are outside the AR core
become much stronger during the second rotation.

The EIS observation sequence for the 19 April 2011 (HPW021 VEL 240x512v1) 
used the 1\arcsec\  slit with an exposure time of 60 seconds. The EIS raster  
started at 12:30:27~UT and finished at 14:34:14~UT. 
The raster was 240\arcsec\ wide and used a slit length of 512\arcsec.
For the second rotation of the active region NOAA~11193 we have chosen 
a full spectral atlas observation made on 17 May 2011.
The EIS raster  started at 00:48  UT and finished one hour later,
at 01:50 UT. We designed this study to have a good S/N, using the 
2\arcsec\ slit,  an exposure time of 60 seconds, and the full 
spectral range of EIS.
We used custom-written software for the EIS data analysis.
For details see \citet{delzanna:2013}.

Fig.~\ref{aia_eis_19_apr} shows a selection of AIA and EIS line 
intensity images, in the core of the AR, with profiles 
of the intensities across the AR on Apr 19, 2011. 
The AIA images are 1-minute averages at 13:30 UT.
Fig.~\ref{aia_eis_17_may} shows similar images for the 
17 May 2011 observation, i.e. 1-minute averages at 01:06 UT.
The AIA images, thanks to their higher spatial and temporal 
resolution, show a much higher spatial variability.

To perform the $DEM$ analysis for the first rotation
 we selected a region, indicated by the vertical lines 
in Fig.~\ref{aia_eis_19_apr} (solar Y=360), 
where we avoided, as much as possible, background 
contamination from the bright moss regions. 
The intensities in the EIS lines formed around 3~MK (e.g. \ion{Fe}{xvi})
are much stronger in the core than in the surroundings, 
so are not affected by background contamination.
There is, however, significant intensity in lines formed around 1~MK
(e.g. \ion{Fe}{ix}). The higher-resolution AIA 171~\AA\ image 
clearly shows that some of the unresolved low temperature emission 
observed by EIS is due to the lower resolution of the spectrometer. 
Some low temperature (1~MK) 
emission is however also present in the AIA images, as found in other 
AR observations \citep{delzanna:2013}.
Whether this 1~MK emission is physically related (in terms of 
nanoflare storm) with  the  3~MK emission is an open question that is not easy to answer.
Estimating an appropriate background is also quite difficult,
so we are providing the results without subtracting  a background.
Clearly,  any slope obtained by EIS 
in the 1--3 MK range is bound to be a lower limit to the actual slope.

For the second rotation (see  Fig.~\ref{aia_eis_17_may}), we selected again a 
region where the  3~MK emission (e.g. \ion{Fe}{xvi}) is strongest.
Moss regions  (strong in \ion{Fe}{xii} and AIA 193~\AA) were avoided.
The intensity in EIS lines formed around 1~MK (e.g. \ion{Fe}{ix}) is low,
but similar to that of nearby regions. 
The AIA 171~\AA\ image clearly  shows this time that 
this  1~MK emission is originating from underlying  cool loops 
that connect the two moss regions.
Again, it is not clear if this cool emission is at all related 
to the 3~MK emission, so also in this case 
we are providing the results without subtracting  a background, and 
the slope obtained from the EIS lines is bound to be a lower limit to the actual one.

\section{Emission measure methods}\label{methods}

We recall that the intensity of an optically thin line 
can be expressed as an integral along the line of sight 
\begin{equation}
I(\lambda_{ji}) =  \int_h N_{\rm e}\, N_{\rm H}\, Ab\, G(N_{\rm e},T_{\rm e})\, dh    
\label{eq:intensity}
\end{equation}
where $Ab$ is the elemental abundance, and $G(N_{\rm e},T_{\rm e})$ is the 
\emph{contribution function} of the  
spectral line. $G(N_{\rm e},T_{\rm e})$ has a very strong dependence on temperature,
 and  a negligible  dependence on the  electron number density, $N_{\rm e}$, 
for a certain range of densities, if the appropriate line is chosen.
Considering only such lines, and assuming that a 
 unique relationship exists between $N_{\rm e}$  and $T_{\rm e}$,
we define 

\begin{equation}
DEM (T) = ~N_{\rm e} N_{\rm H} \frac{dh}{dT} \quad~~~[cm^{-5} K^{-1}]
\end{equation}
as the column differential emission measure ($DEM$) of the plasma, which
gives an indication of the amount of plasma along the line of 
sight that is emitting the radiation observed at a temperature between 
$T$ and $T+dT$. The $DEM$  by definition is a continuous distribution
in a range of temperatures.

The $DEM$ inversion is an ill-posed problem with various complexities
associated, see e.g. 
 \citep{craig_brown:76,craig_brown:86,judge_etal:97,delzanna_thesis:99,delzanna_etal:02_aumic}
and references therein for details.

We have seen in previous studies that the methods of the inversion 
normally provide similar results, if the DEM is relatively well constrained 
and the same input parameters/temperature range is adopted 
(see, e.g. \citealt{delzanna_etal:11_aia,odwyer_etal:2014,delzanna_mason:2014}).
However, to assess the sensitivity of the present results to the method adopted, we have  run three different 
DEM methods:   the spline method described in \citet{delzanna_thesis:99}; 
the Markov Chain Monte Carlo method of \citet{KasD:98} (MCMC\_DEM);
and the XRT\_DEM\_ITERATIVE2 method, available within SolarSoft 
(see \citealp{weber_etal:2004}). The last two methods are widely used in the literature.

We also consider the column emission measure $EM(0.1)$, calculated by 
integrating the $DEM$ over the temperature bins $\Delta$ log $T$= 0.1,
to estimate the slope.
There are various EM approximations used in the literature
 (see \citealp{delzanna_thesis:99} for details).
Following \cite{pottasch:63}, many authors 
have approximated the intensity of a line 
\begin{equation}
I(\lambda_{ji})  =   Ab ~ <G(T)> ~ {\int_h  ~N_{\rm e} N_{\rm H} dh }
\end{equation}
by  estimating an averaged value of G(T). The method assumes that each line is 
mainly formed at temperatures close to the peak value $T_{\rm max}$ of its 
contribution function.  \cite{pottasch:63} adopted $<G(T)> = 0.7 ~C(T_{\rm max})$.
A slightly different approximation was suggested by 
\cite{jordan_wilson:1971}.
\cite{jordan_wilson:1971} assumed  that $G(T)$ has a 
constant value over a narrow  interval centred around 
  the temperature of maximum ion abundance in ionisation equilibrium.
Here, we adopt a slight modification, by using 
 the temperature  $T_{\rm max}$  at which
$G_{\lambda}$ has its maximum. 
This approximation was used by e.g. \cite{TriMD:10} to estimate the 
EM slopes.

Following \citet{jordan_wilson:1971}  we define
\begin{equation}
G_{\lambda,0}(T)= \left\{ \begin{array}{ll}
 C_{\lambda} & |{\rm log}\,T - {\rm log}\,T_{\rm max}| < 0.15 \\
 \\
           0 &\quad |{\rm log}\,T - {\rm log}\,T_{\rm max}|  > 0.15\\
 \end{array}
                         \right. 
\end{equation}
and require 
\begin{equation}
\int G_{\lambda}(T) dT = \int G_{\lambda,0}(T) dT
\end{equation}
so that 

\begin{equation}\label{em2}
G_0 = \frac{\int G(T_{e}, N_{e})~dT_{e}} {T_{\rm max} \times (10^{0.15} - 10^{-0.15})}.
\end{equation}

The values of the constant $C_{\lambda}$ are calculated using the CHIANTI
routine {\sc integral\_calc}.
The estimate for the emission measure at the 
temperature of  a single line, $EM_{\rm jw}$, is then simply obtained
from the observed intensity $I_{\rm o}$:
\begin{equation}
EM_{\rm jw}= \frac{I_{\rm o}} {Ab~ C_{\lambda} }
\end{equation}

The points are normally close to the minima of the EM loci 
curves, if the emission lines are not blended.
The minima of the EM loci curves  ($I_{\rm o}/G_{\lambda}(T)$) 
are by definition upper limits to the EM distribution, which 
however can be very different
 (see  \cite{delzanna_thesis:99,delzanna_etal:02_aumic} for details).

\begin{table}[!htdp]
\centering
\caption{Spectral lines used for the DEM inversion.} 
\begin{tabular}{llll}
\hline
Ion & $\lambda$ (\AA) & $\log\, T_{\rm max}$ [K]  \\
\hline
Fe~VIII (bl) & 185.2   	&5.70   \\
Fe~VIII (bl) & 186.6   	&5.70   \\
Si~VII & 275.4   	&5.80   \\
Fe~IX       &  197.9   	&5.95  \\
Fe~IX       &  188.5   	&5.95  \\
Fe~X        & 184.5  	&6.05  \\
Fe~XI      &  180.4  	&6.15  \\
Fe~XI      &  188.2  	&6.15  \\
Fe~XII    &192.4  	&6.20  \\
Fe~XII    & 195.1  	&6.20  \\
Fe~XIII & 202.0  	&6.25  \\
Fe~XIV & 264.8  	&6.30  \\
Fe~XV  & 284.2  	&6.35  \\
Fe~XVI & 263.0  	&6.45  \\
Ca~XIV & 193.9  	&6.55  \\
Ca~XV & 201.0   	&6.6   \\
Ca~XVI & 208.5  	&6.7   \\
Fe~XVII & 254.9  	&6.75   \\
\hline
\end{tabular}
\end{table}


We used CHIANTI v.7.1 \citep{landi_etal:12_chianti_v7.1}
atomic data and ionization equilibrium tables.
We use the new set of abundances for the 
3~MK loops  obtained by \citet{delzanna:2013,delzanna_mason:2014}.
We note that  these abundances present a first ionisation potential (FIP)
 enhancement of about a factor of 3.2, slightly lower than 
the factor of 4 of the \citet{Fel:92} coronal abundances.
The fact that active region cores are better represented by coronal abundances
was already pointed out by \citet{TriKM:11}.
However, we point out that our DEM results  between 1 and 3 MK are entirely 
independent of elemental abundance issues, since they are based only on iron lines.
The list of the spectral lines used for DEM analysis is given in Table~1.
They are mostly from iron, although a few low FIP elements were also used to constrain 
lower and higher temperatures (outside the range 1-3MK).

\section{Results}

\subsection{Spatial variability}

 \begin{figure*}[!htbp]
\centerline{
 \epsfig{file=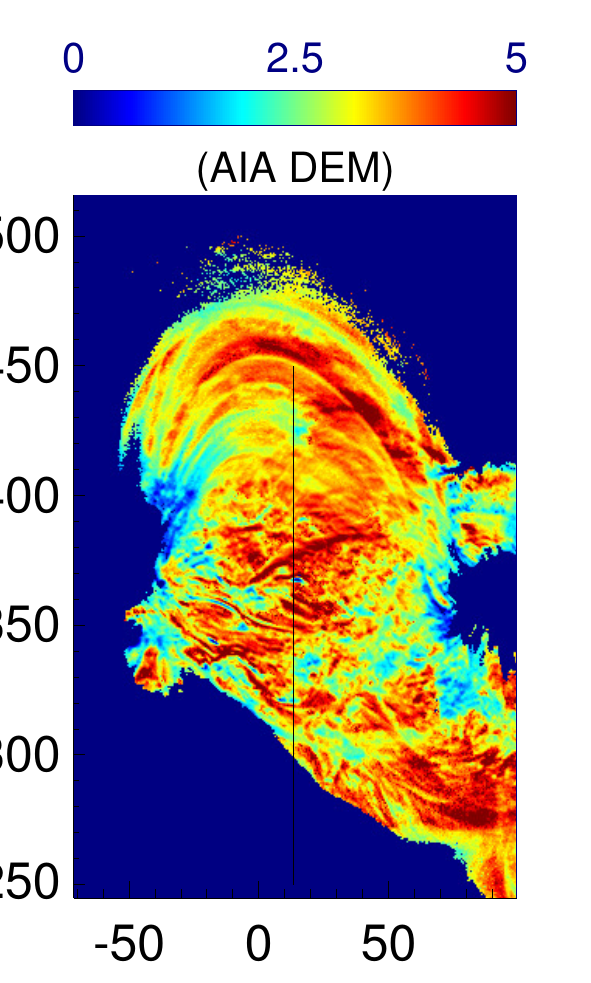, width=2.6cm,angle=0 }
 \epsfig{file=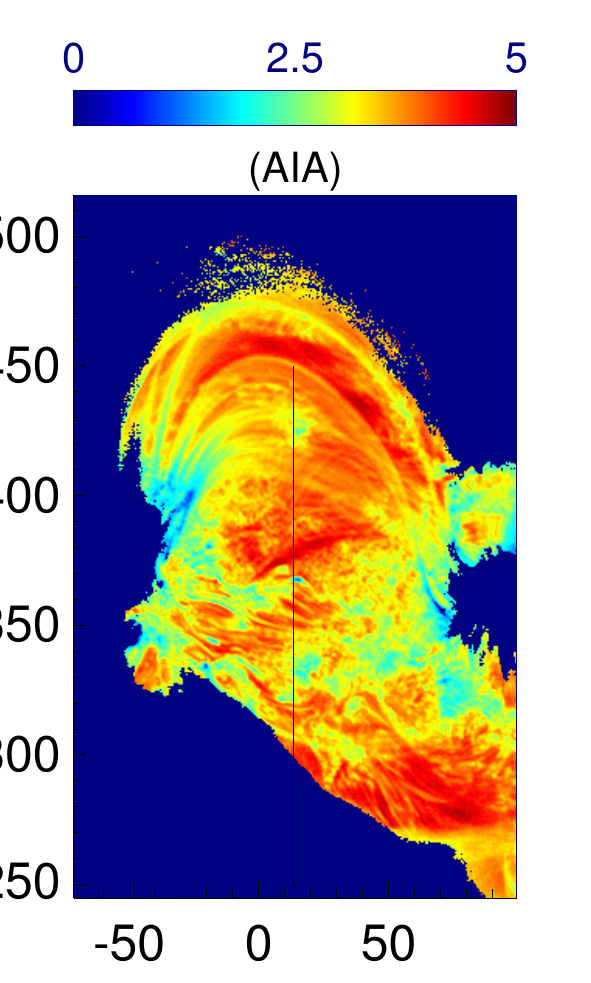, width=2.6cm,angle=0 }
 \epsfig{file=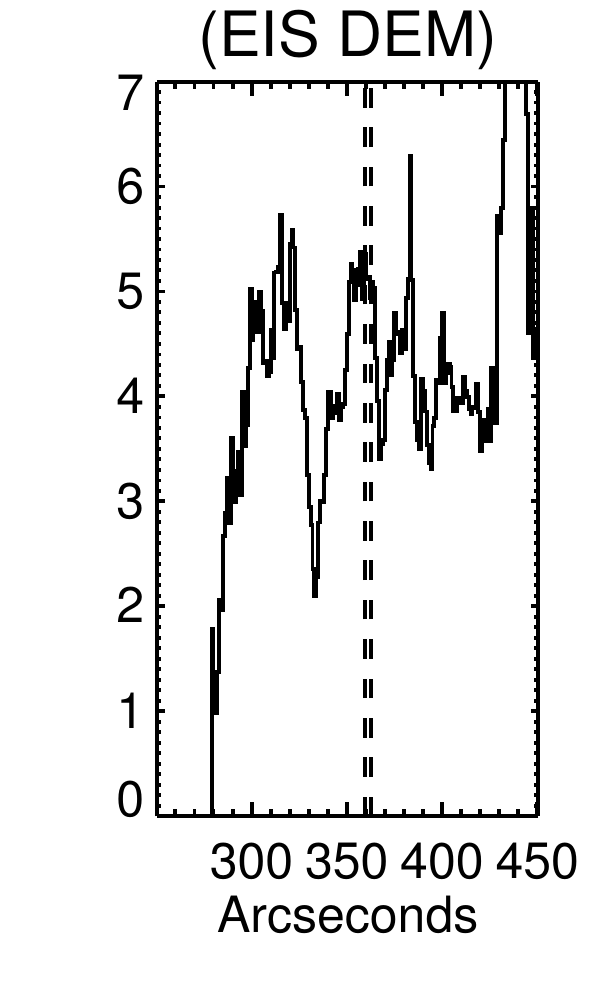, width=2.6cm,angle=0 }
 \epsfig{file=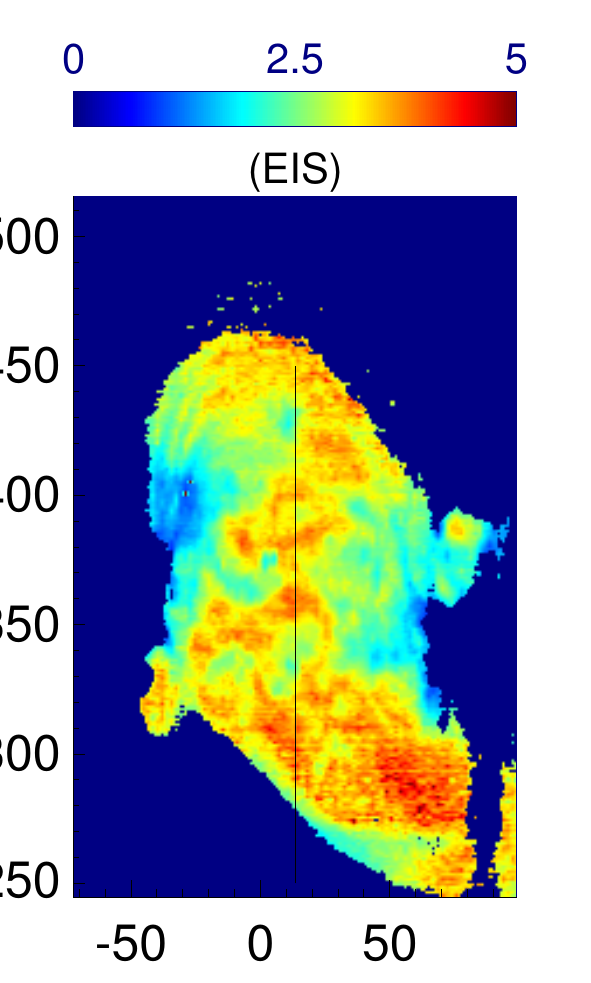, width=2.6cm,angle=0 }
}
 \centerline{
 \epsfig{file=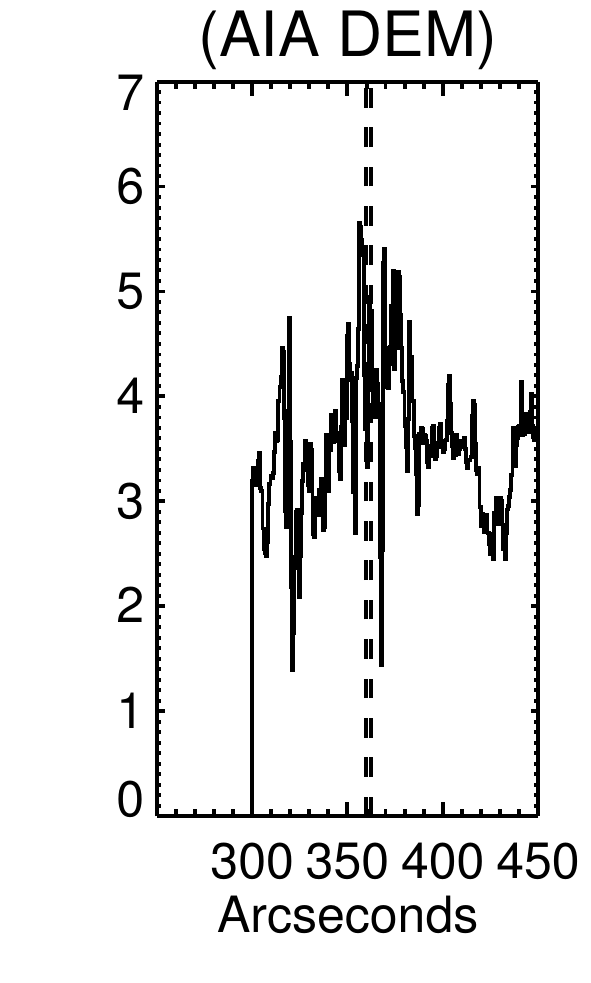, width=2.5cm,angle=0 }
 \epsfig{file=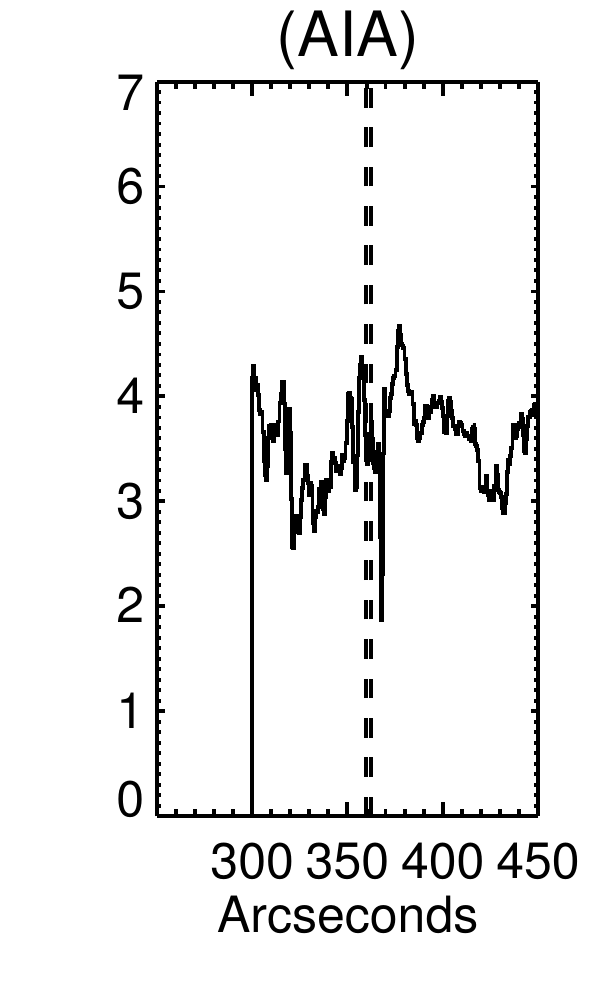, width=2.5cm,angle=0 }
 \epsfig{file=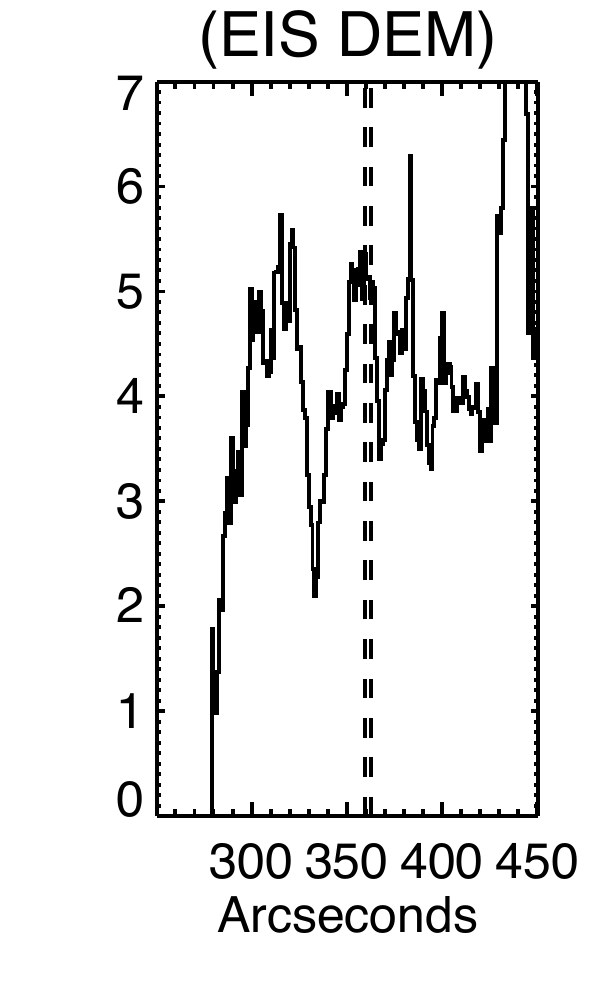, width=2.5cm,angle=0 }
 \epsfig{file=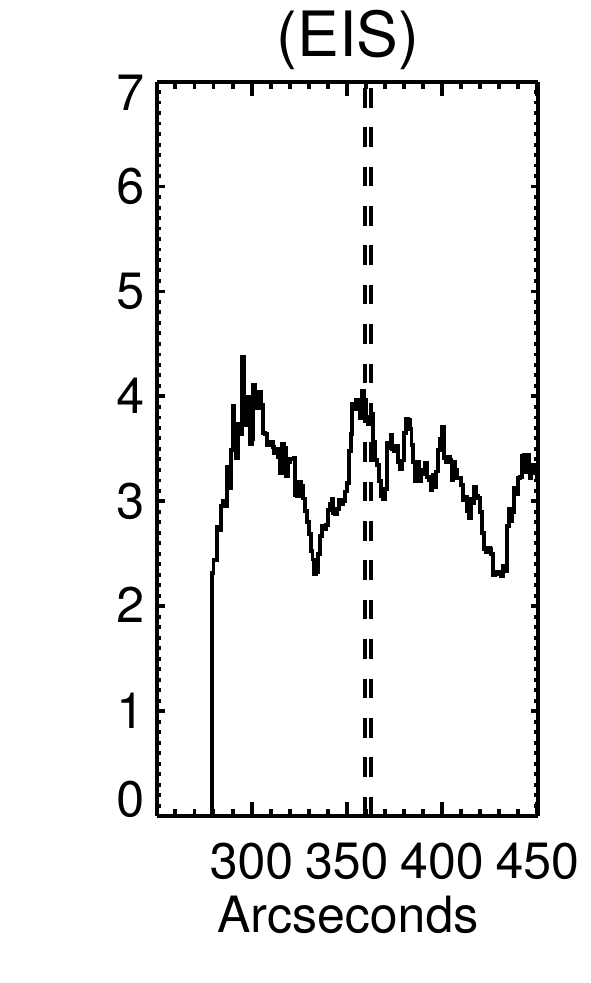, width=2.5cm,angle=0 }
}
 \centerline{
 \epsfig{file=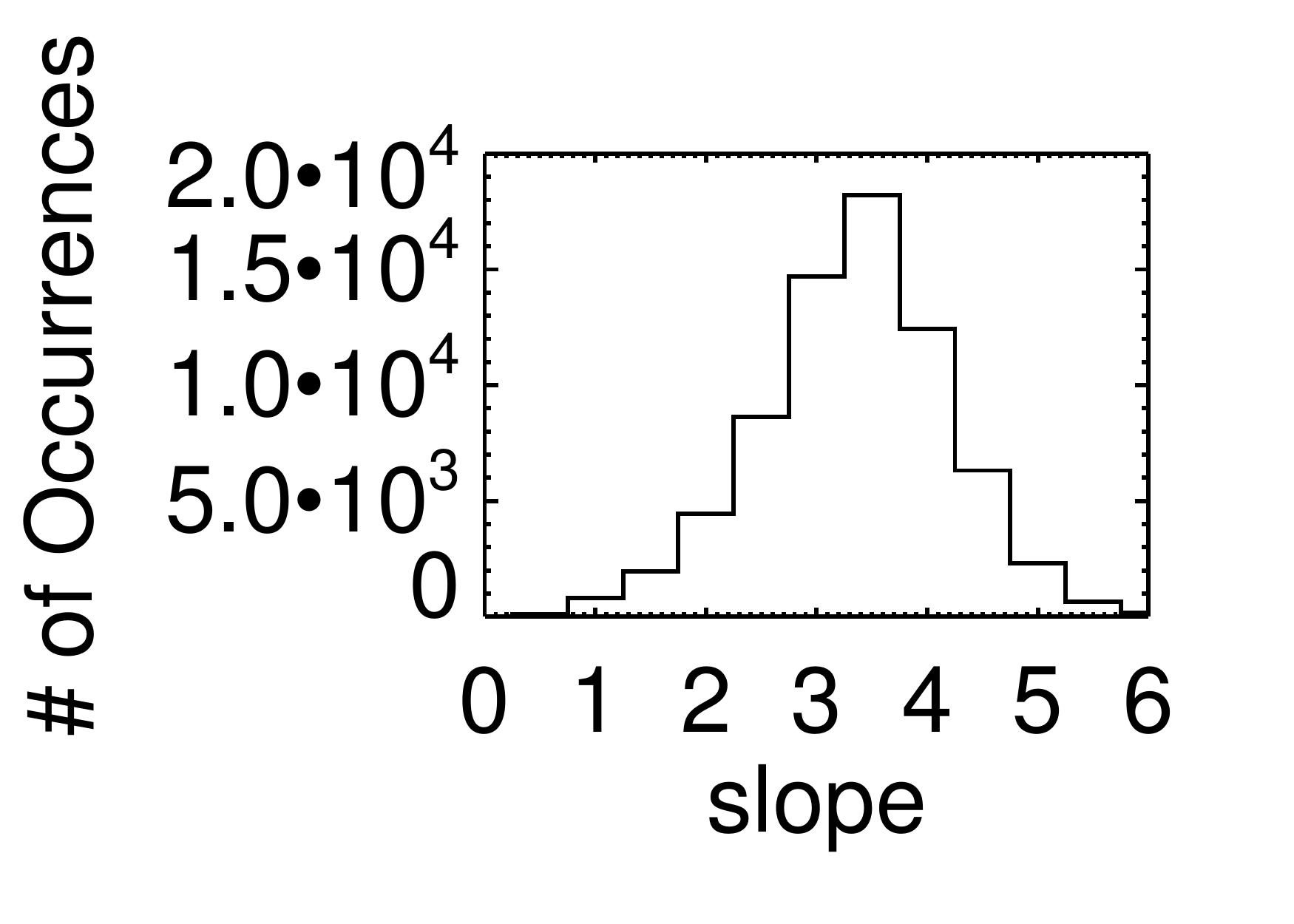, width=2.5cm,angle=0 }
 \epsfig{file=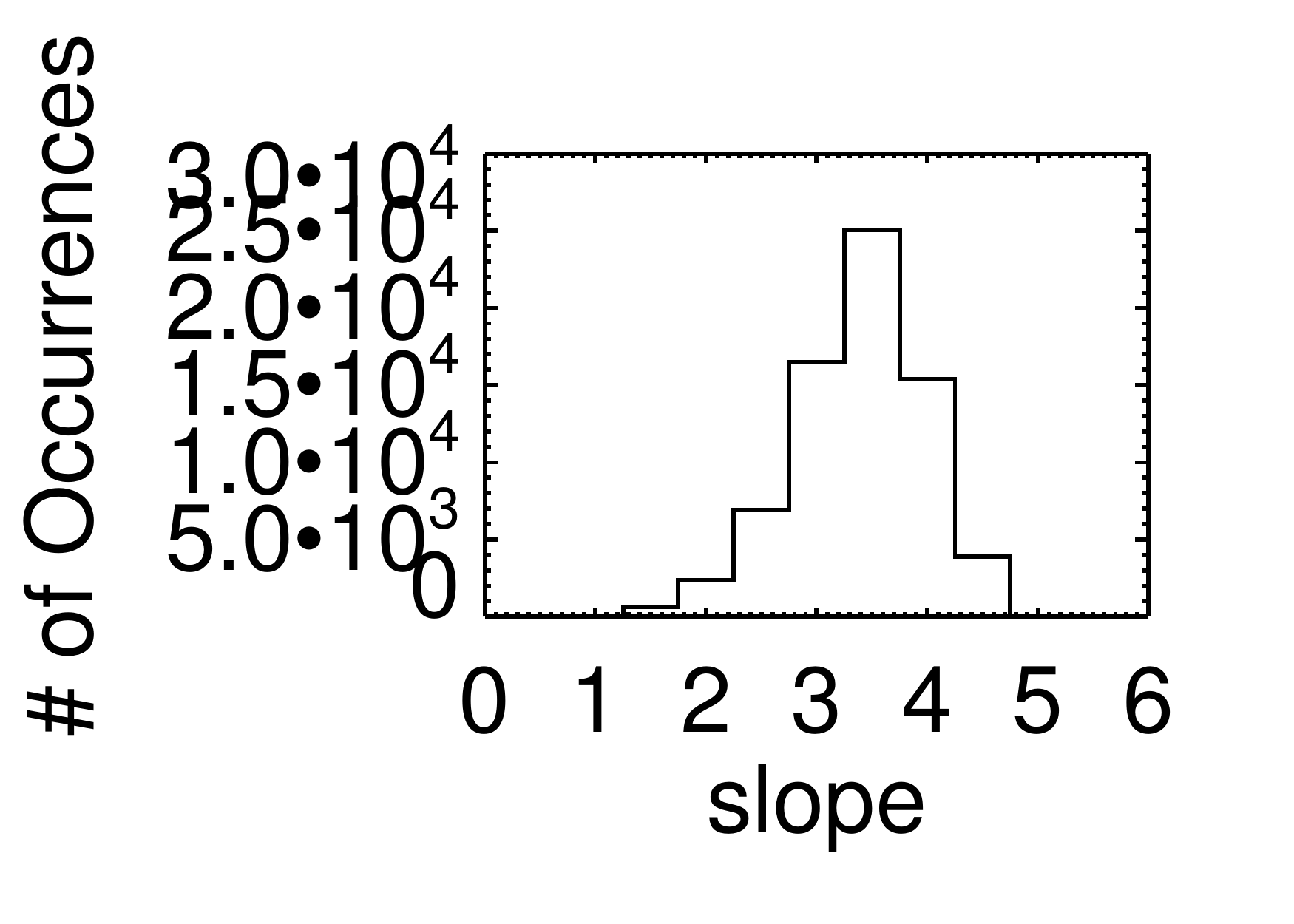, width=2.5cm,angle=0 }
 \epsfig{file=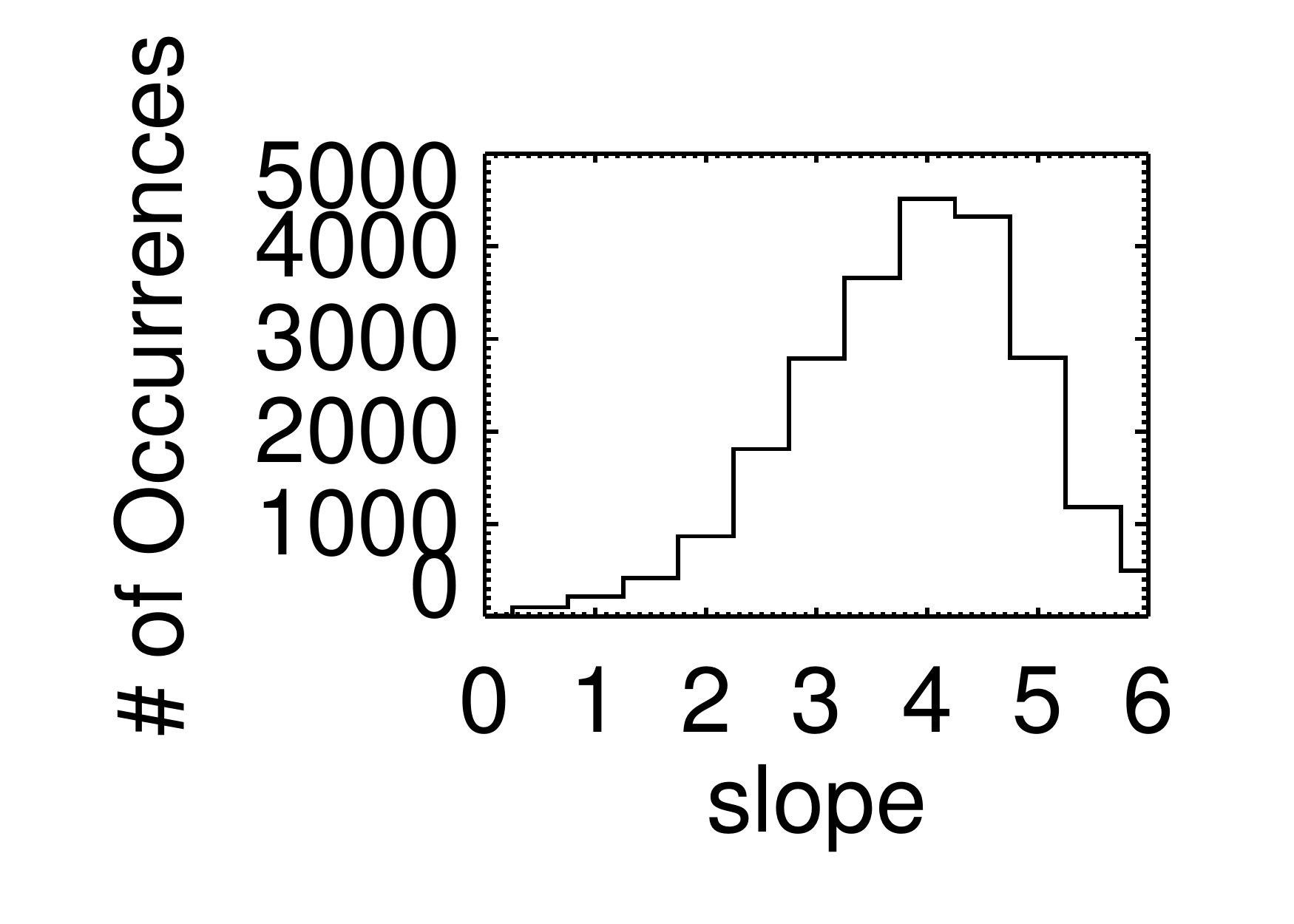, width=2.5cm,angle=0 }
 \epsfig{file=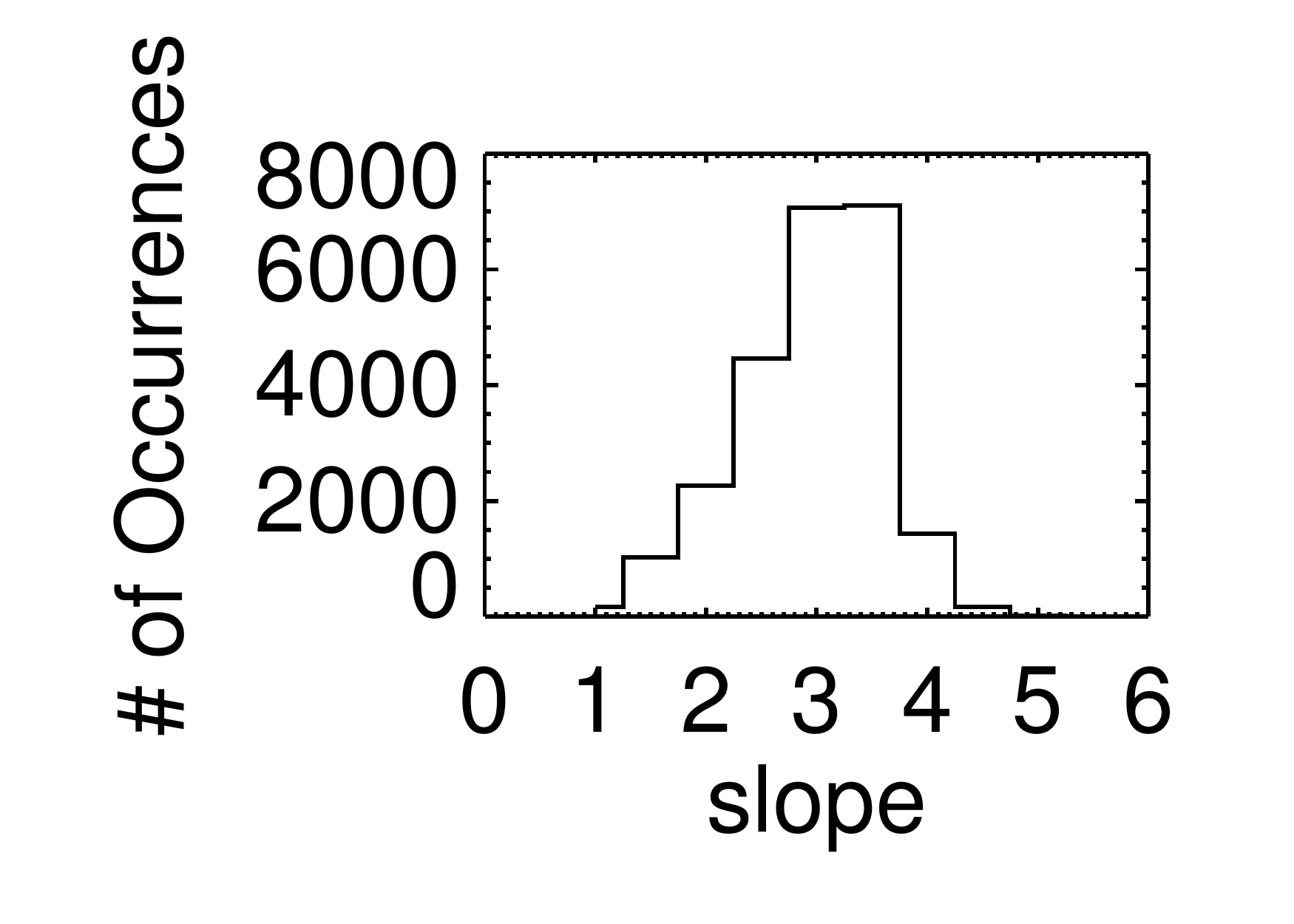, width=2.5cm,angle=0 }
}
 \caption{The slopes of the EM distribution in the 1--3 MK temperature range,
 as estimated from the AIA DEM, the approximate Pottasch method applied to the 
 AIA data, the EIS DEM,  and the Pottasch method applied to the EIS observations.
 The AIA slopes are obtained from  full-resolution 
(1\arcsec) images averaged over 13:30--13:31 UT.
The top row shows the images of the slopes, the middle the profiles along the 
vertical line shown in the images, and the bottom row the  histograms of the 
distribution of the slopes.
  }
 \label{fig:slopes_1}
 \end{figure*}

 \begin{figure*}[!htbp]
 \sidecaption
 \centerline{
\epsfig{file=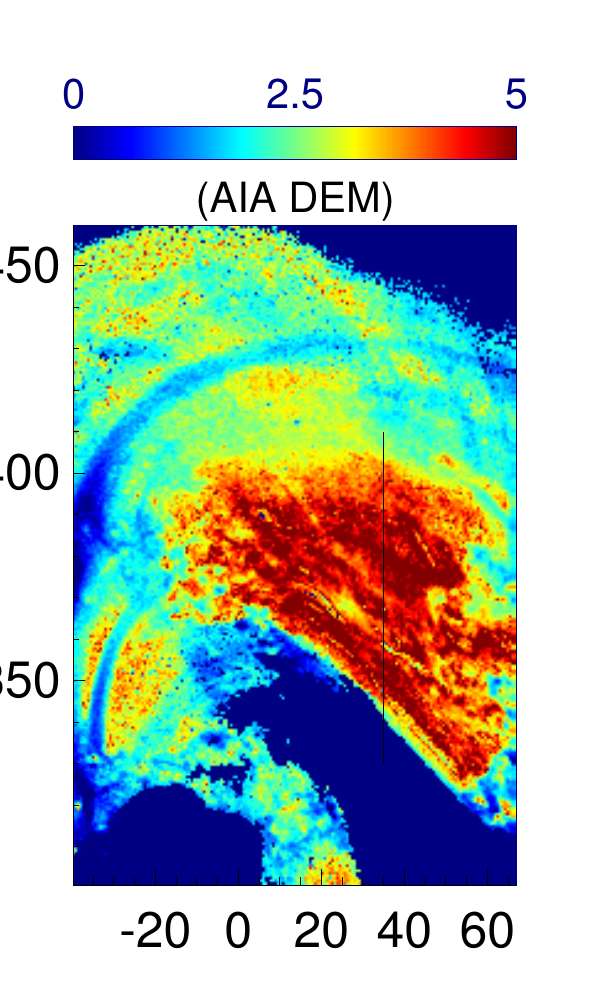, width=2.6cm,angle=0 }  
\epsfig{file=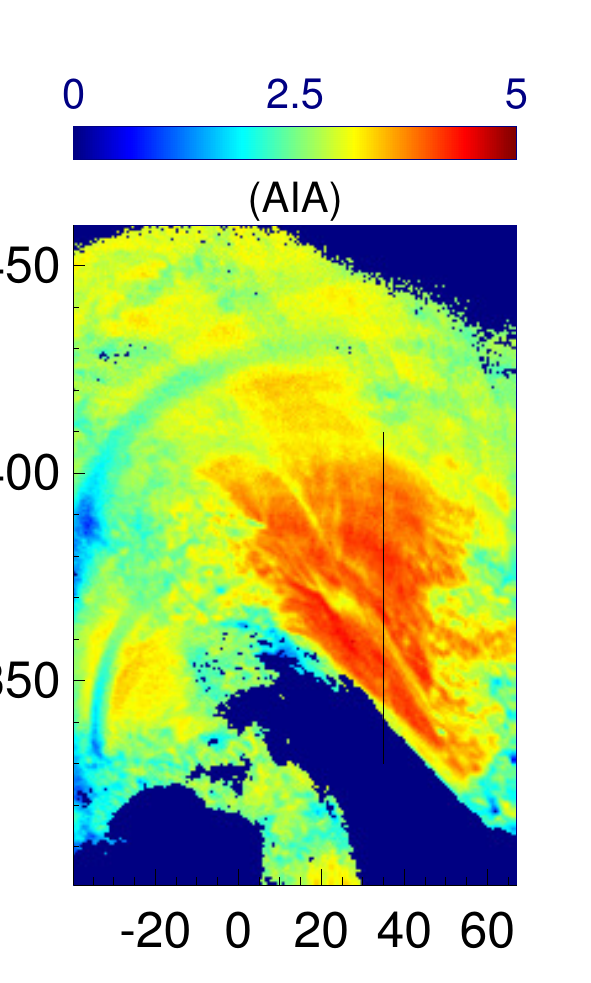, width=2.6cm,angle=0 }
\epsfig{file=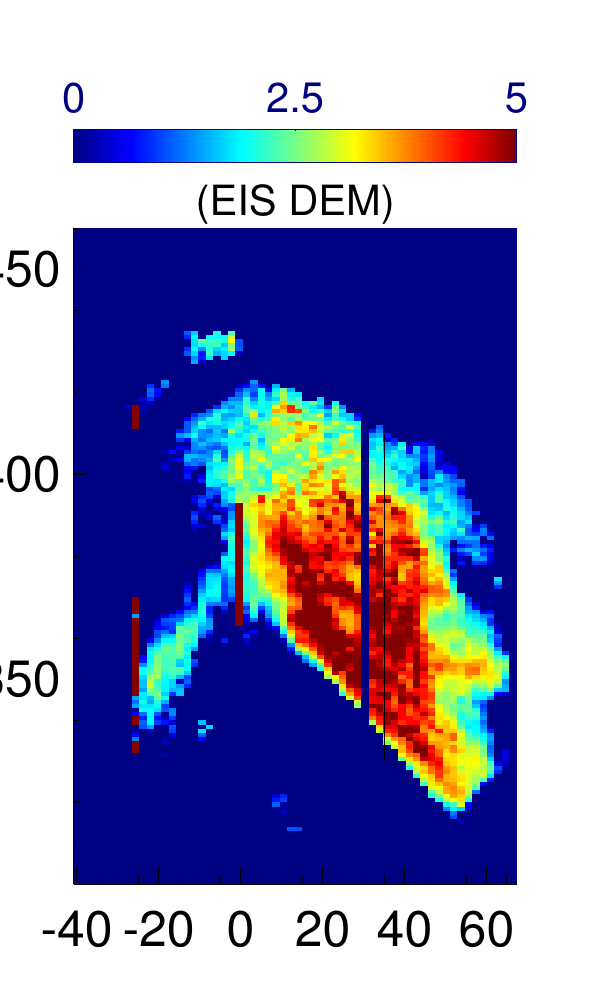, width=2.6cm,angle=0 }
\epsfig{file=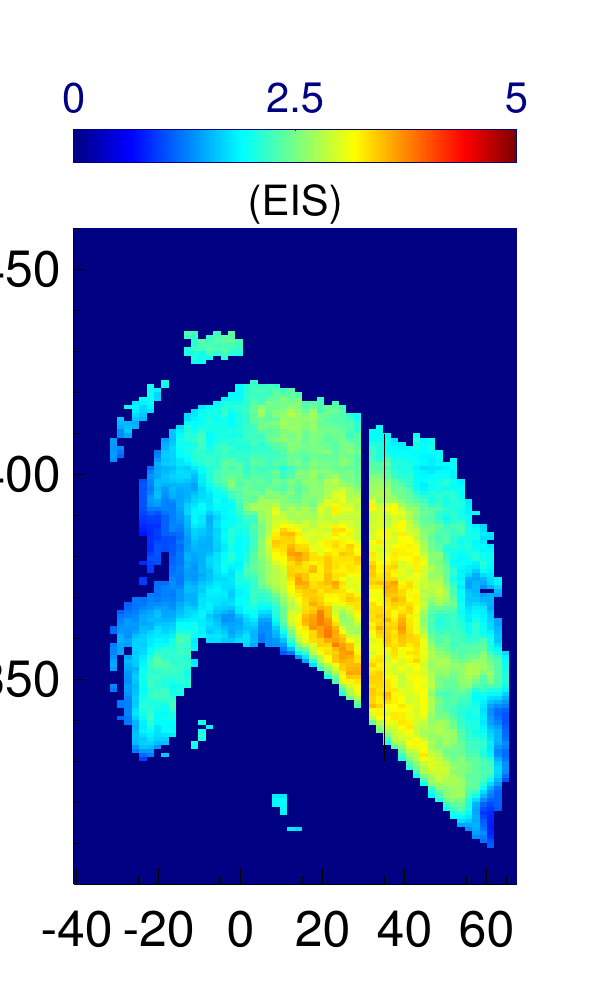, width=2.6cm,angle=0 }
}
 \centerline{
 \epsfig{file=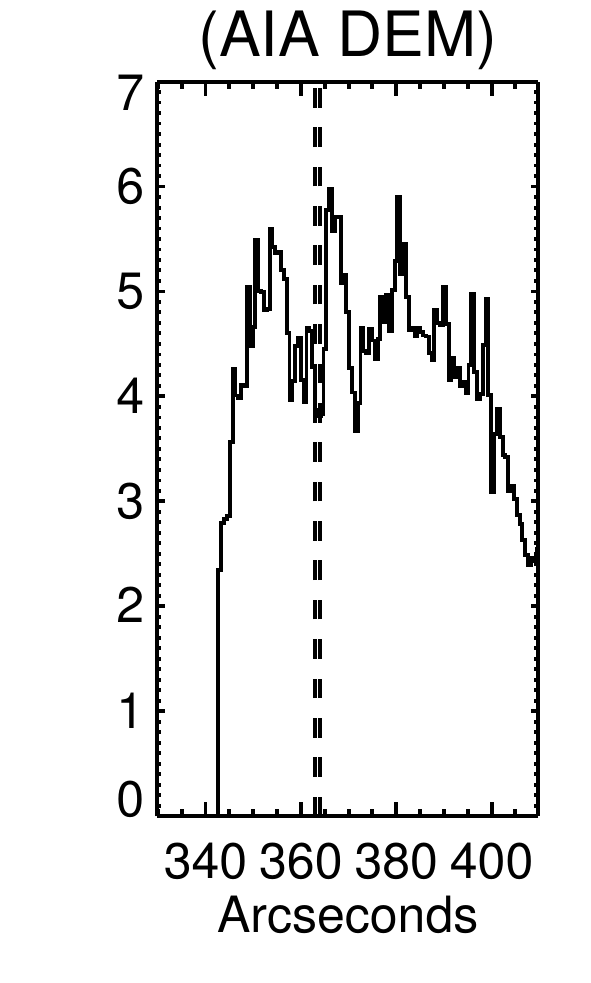, width=2.5cm,angle=0 }
 \epsfig{file=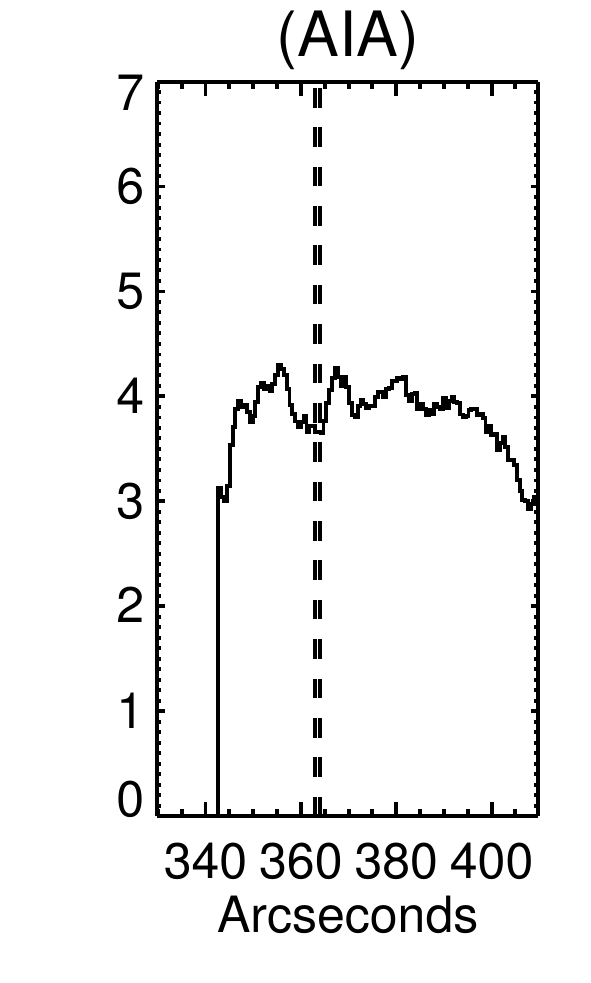, width=2.5cm,angle=0 }
\epsfig{file=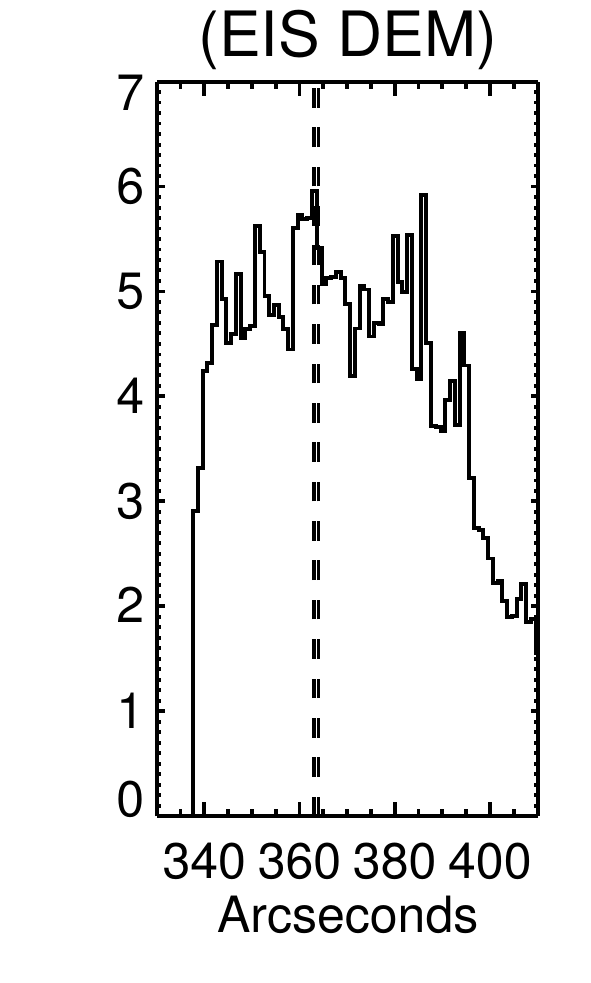, width=2.6cm,angle=0 }
\epsfig{file=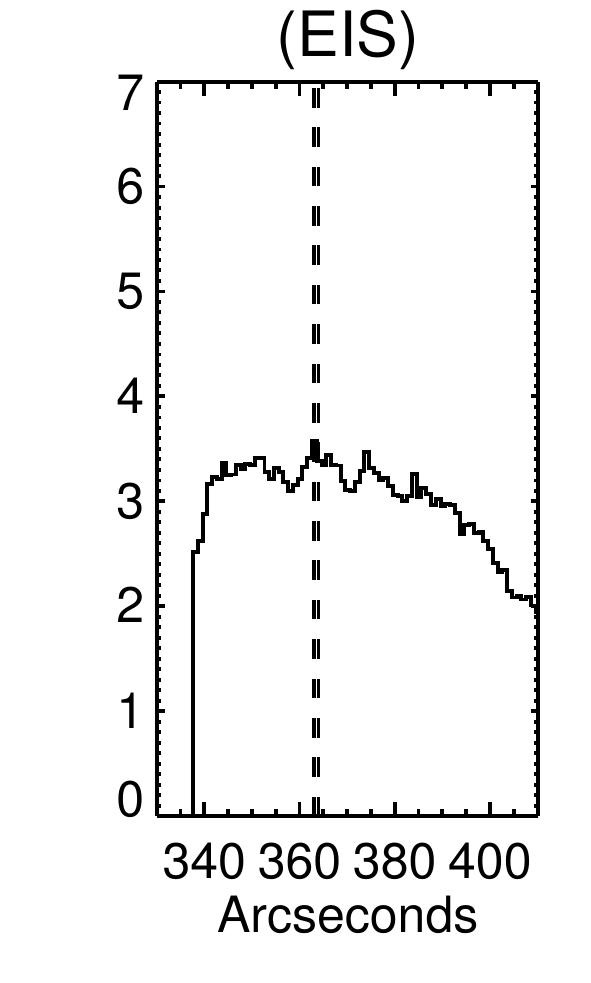, width=2.5cm,angle=0 }
}
 \centerline{
 \epsfig{file=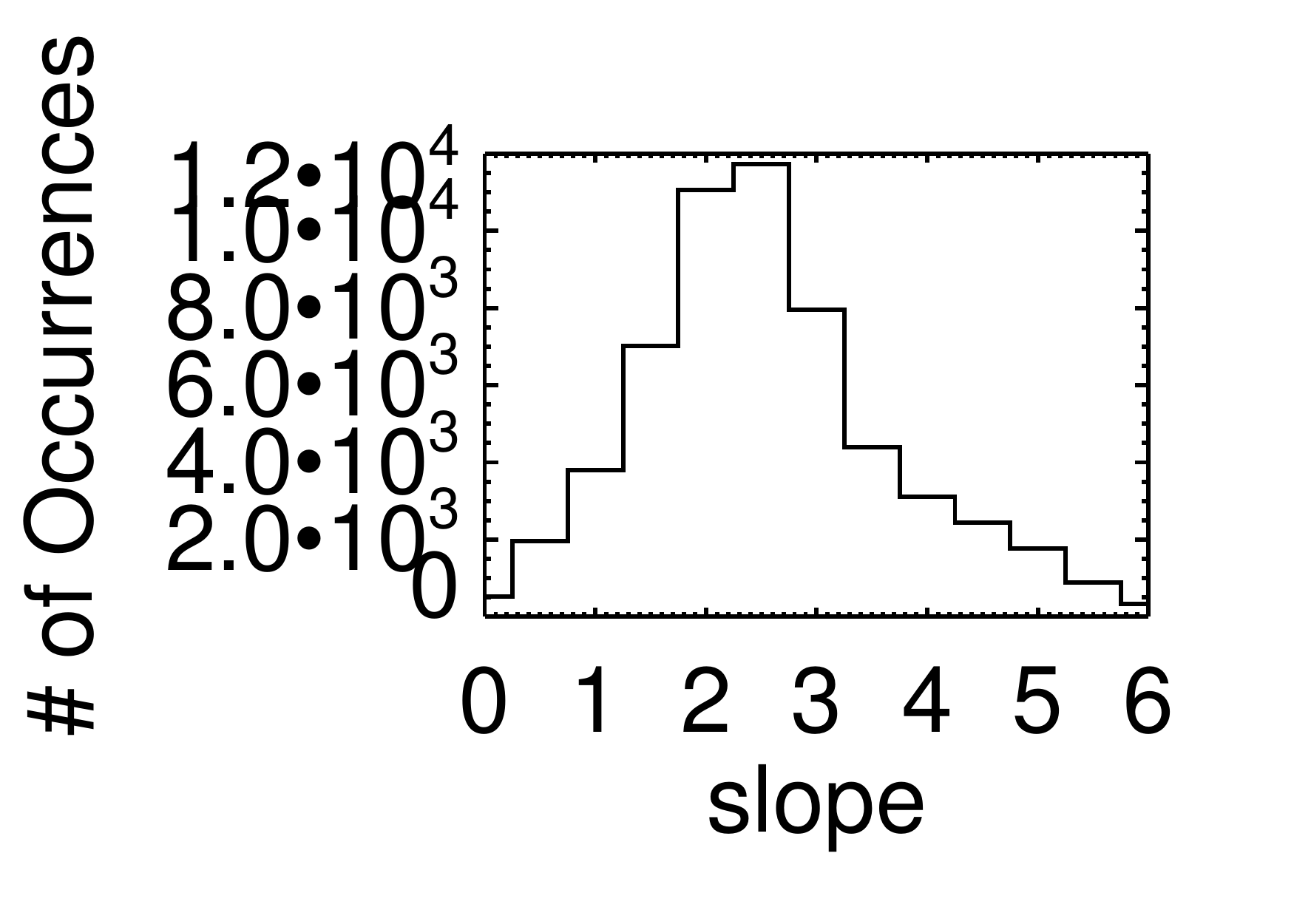, width=2.5cm,angle=0 }
 \epsfig{file=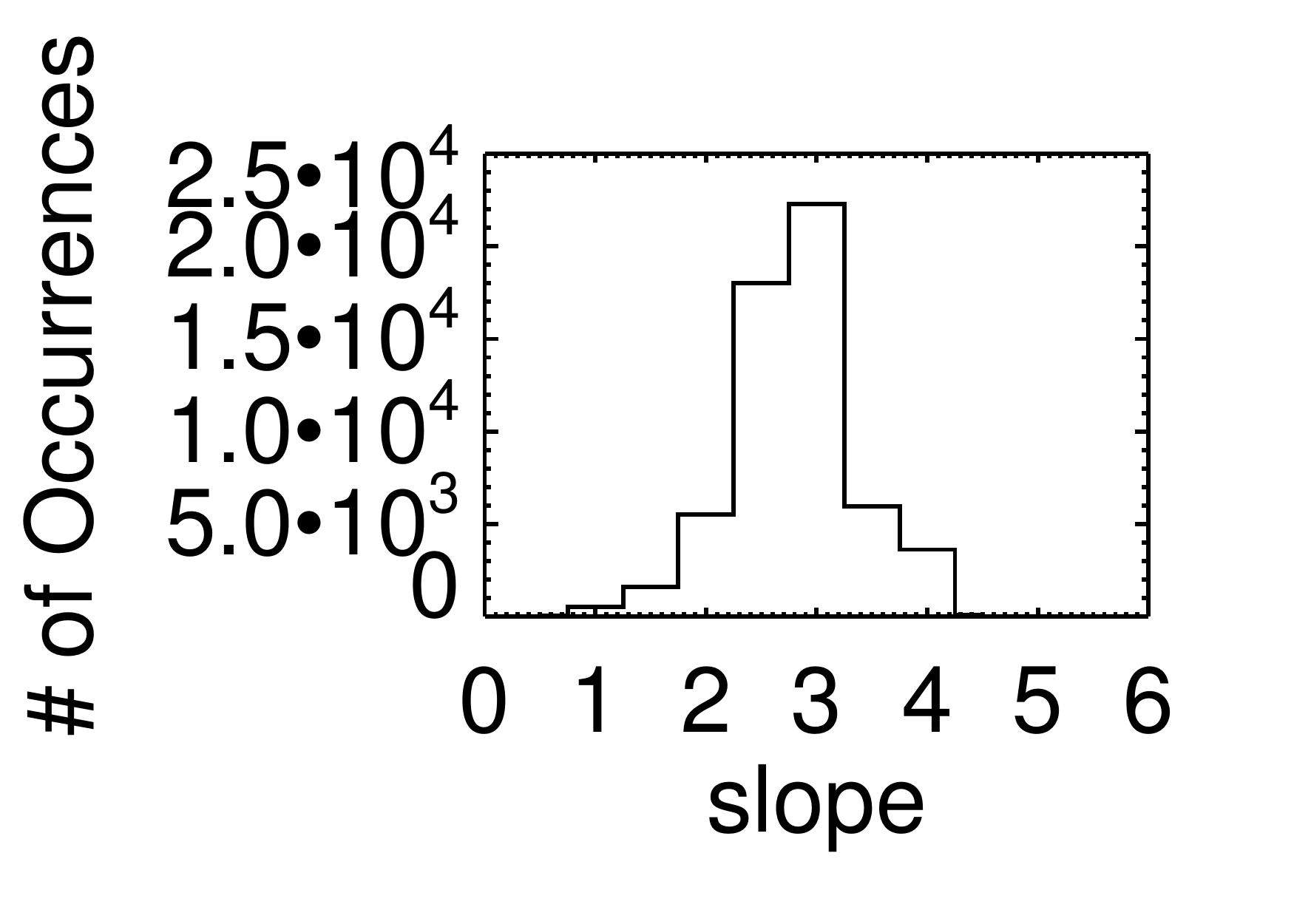, width=2.5cm,angle=0 }
 \epsfig{file=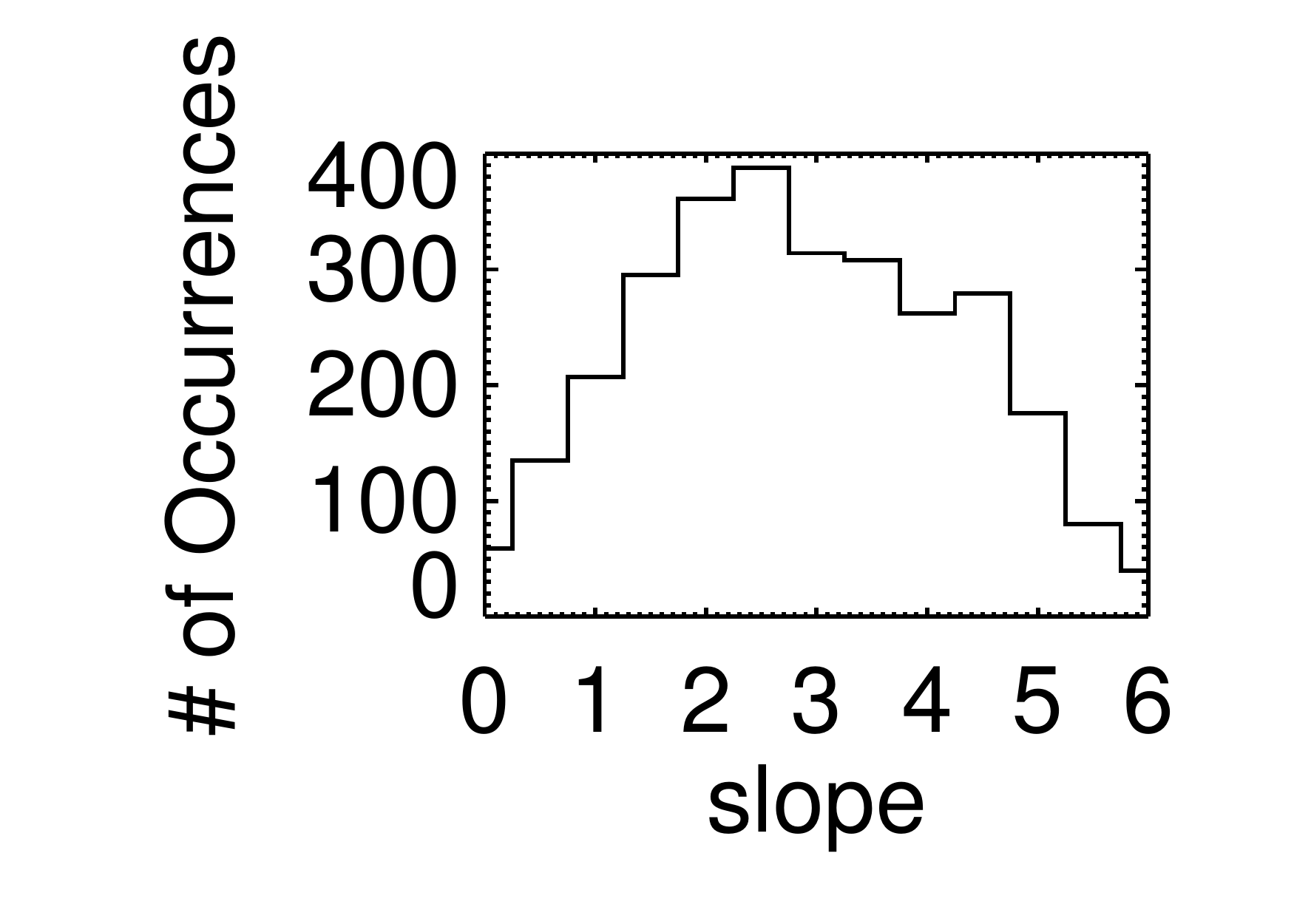, width=2.5cm,angle=0 }
 \epsfig{file=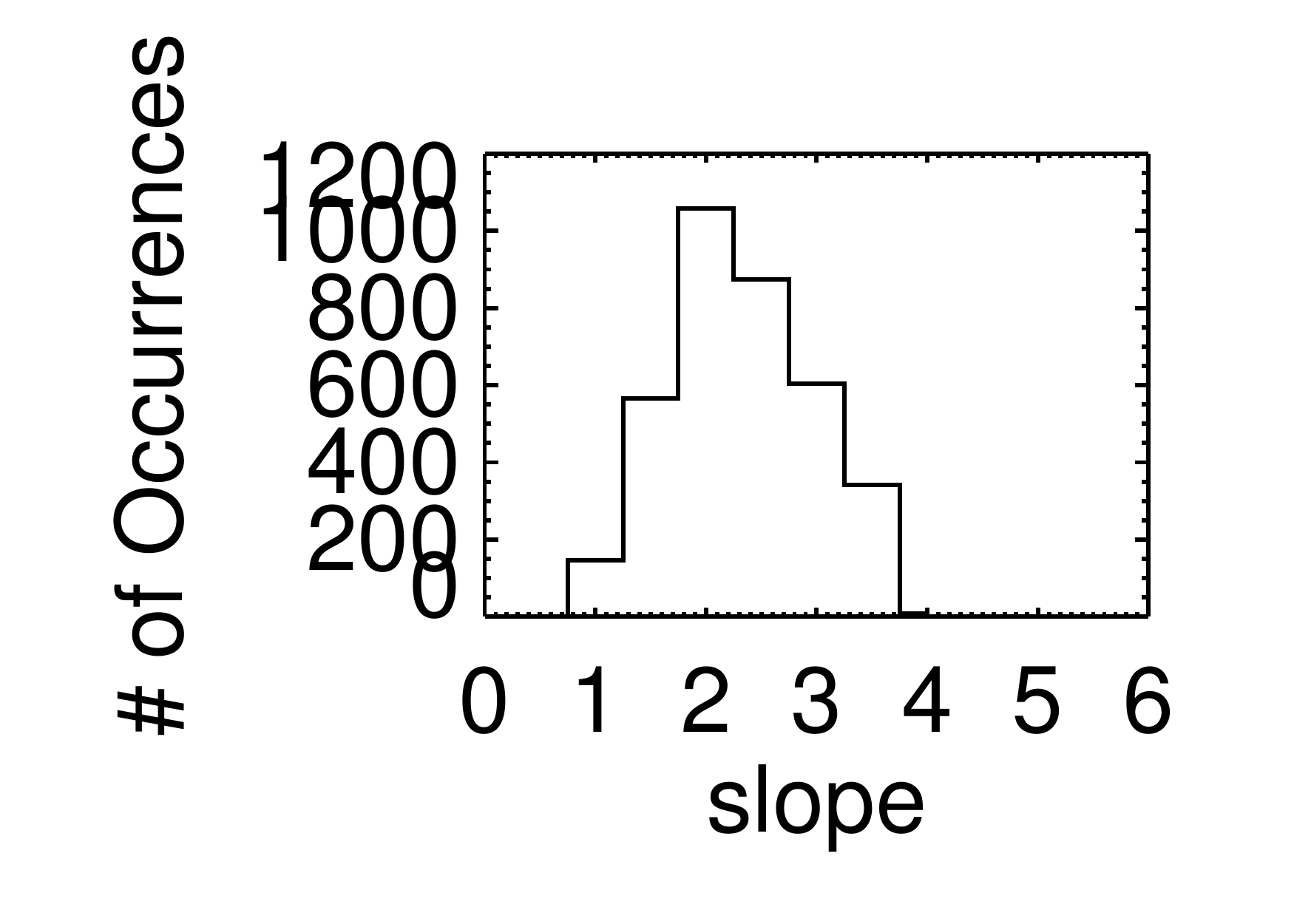, width=2.5cm,angle=0 }
 }
\caption{Same as  Fig.~\ref{fig:slopes_1}, for the second rotation} 
 \label{fig:slopes_2}
 \end{figure*}

One natural question we wanted to address is the amount of 
 spatial variability within an AR. 
Another issue that we wanted to address is whether the lower spatial resolution of 
EIS  had any influence on the result. 
The higher-resolution  AIA images clearly show a lot more loop structures within the 
cores of ARs, so in principle we could expect higher slopes. 
Another question is how much the results depend on the 
different methods. We compare the results of  four  methods in  Fig.~\ref{fig:slopes_1}. 

First, we obtained a DEM for each AIA pixel using a faster version of the 
regularised inversion method described by \cite{hannah_kontar:2012}.
We used as kernel the AIA responses that we calculated using CHIANTI v.7.1 
\citep{landi_etal:12_chianti_v7.1}, and the \cite{delzanna:2013} AR core abundances.
We then fitted a line in the log $T$-log $EM$  plot in the 1--3 MK range,
and obtained the slopes shown in Fig.~\ref{fig:slopes_1} (first column from the left).
The slopes, over the whole of the AR core, range between 2 and 5,
with the most frequent value around 3.5.

Second, we have considered a faster method (similar to the \citealt{pottasch:63} one)
applied to the AIA data.
We have considered the intensity of the 171~\AA\ band as only being due to 
Fe IX, and used the AIA 171~\AA\ response function instead of the $G(T)$ to 
estimate the EM at 1~MK. To estimate the EM at 3~MK, we 
have calculated the AIA 335~\AA\ response function by including only the 
emissivity of the Fe XVI 335~\AA\ line, and estimated the observed 
AIA  DN/s  due to Fe XVI 335~\AA\ following the method developed by \citet{delzanna:2013}. 
The estimate of the EM slope in the 1--3~MK 
range obtained from AIA is  shown in Fig.~\ref{fig:slopes_1}  (second column from the left).
The variation in the slopes is much less, but still with the most frequent value around 3.5.

Third, we obtained the slopes applying the 
XRT\_DEM\_ITERATIVE2 method to the EIS intensities to obtain the DEM.
We then calculated the EM(0.1) values, and fitted them in the 1--3 MK
range to obtain the slopes shown in  Fig.~\ref{fig:slopes_1}  
(third column from the left).
The slopes are slightly higher, with the most frequent value
 around 4.

Finally,  we computed the 
$EM_{\rm jw}$ values for two EIS lines  emitted near 1 (\ion{Fe}{ix} 188.5~\AA) and 
3~MK (\ion{Fe}{xvi} 263.0~\AA) and obtained the slope by fitting
a straight line. The last column  in Fig.~\ref{fig:slopes_1}  shows
the slope obtained in this way. 
We can  see that as in the AIA case, the slopes in the core of the active region
have less variability and are slightly lower, around 3.

We performed the same analysis on the second rotation. The results 
are shown in Fig.~\ref{fig:slopes_2}. 
The slopes in the areas where the \ion{Fe}{xvi} emission is low are masked out
(in blue) in  Figs.~\ref{fig:slopes_1},\ref{fig:slopes_2}. 
They correspond to areas where the DN/s in the 
\ion{Fe}{xvi} 263.0~\AA\ line were below 60, and where the estimated DN/s due to the 
\ion{Fe}{xvi} 335~\AA\ line in the AIA 335~\AA\ channel were below 20.

\subsection{DEM of the AR core for the first rotation}

\begin{figure}[!htbp]
\centerline{\epsfig{file=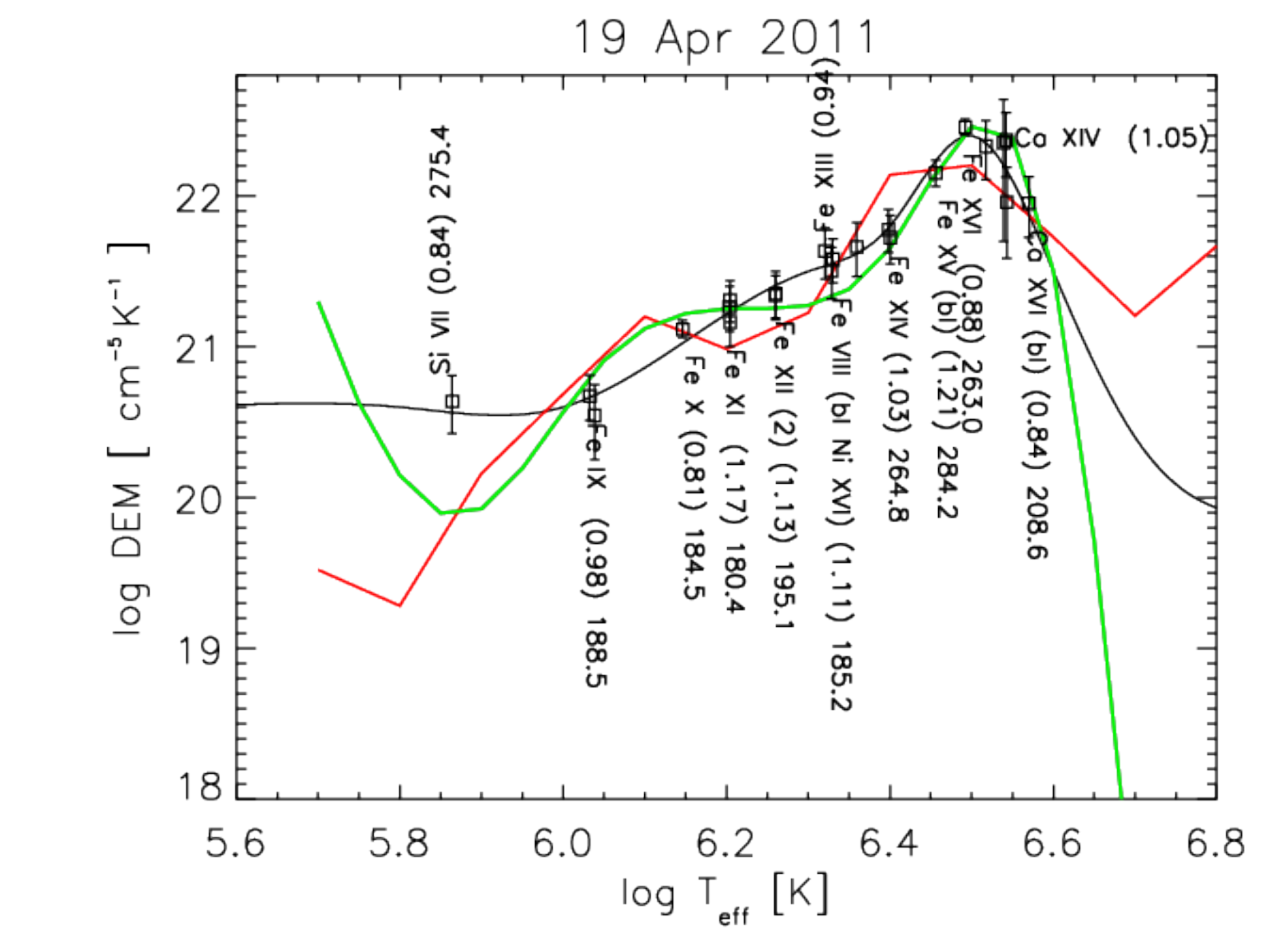, width=6.5cm,angle=90}}
\centerline{\epsfig{file=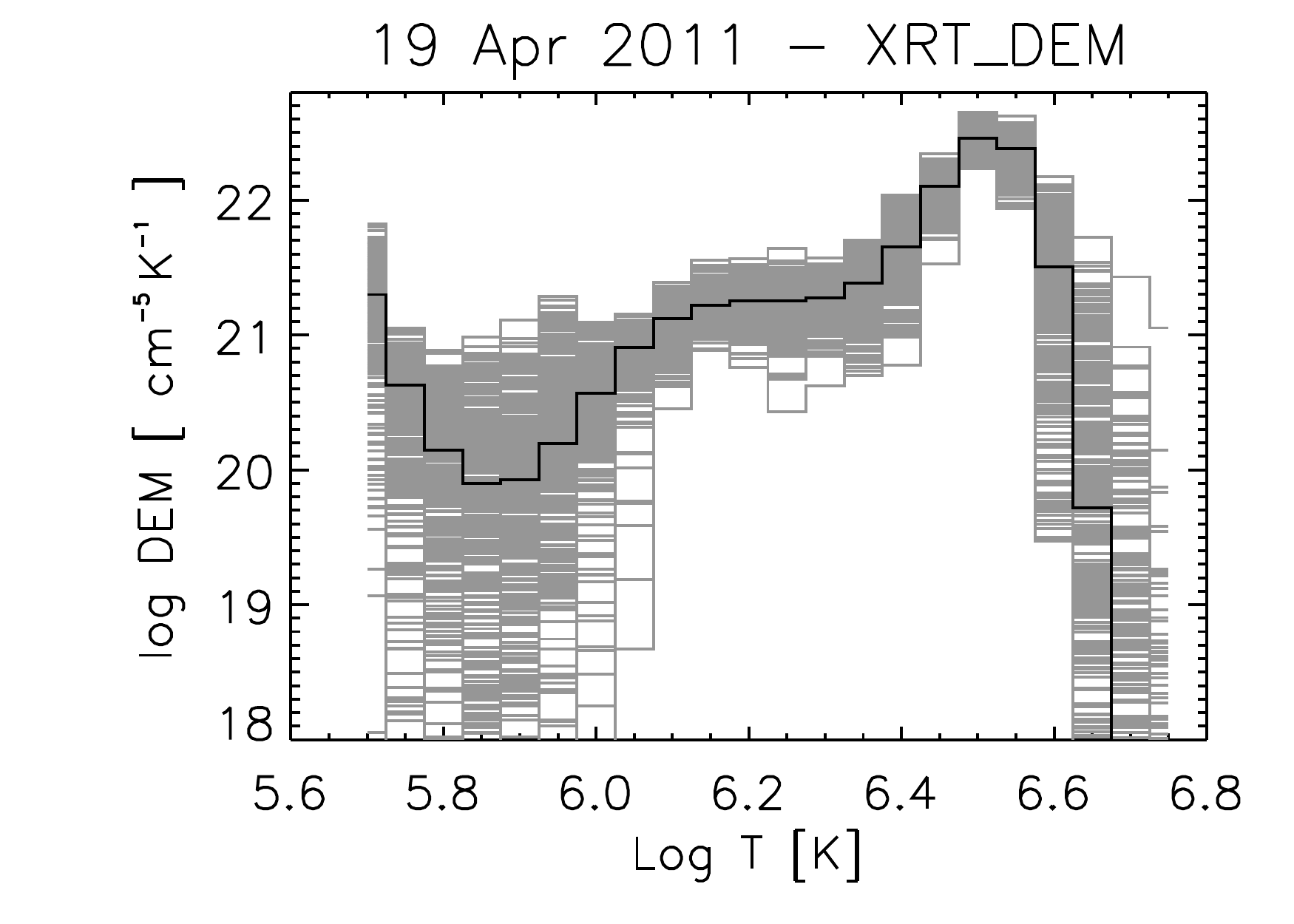, width=6.cm,angle=0}}
\centerline{\epsfig{file=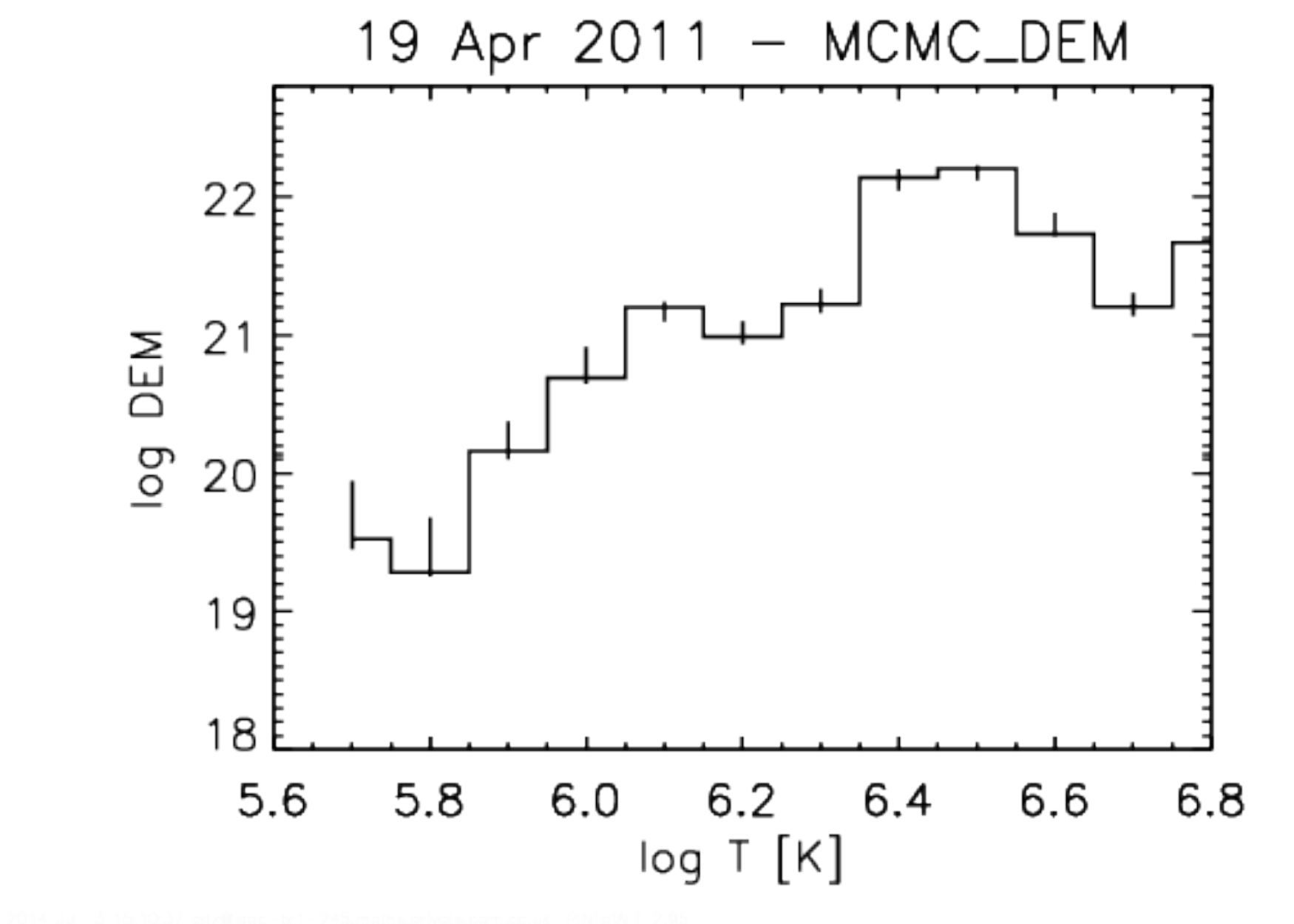, width=6.cm,angle=0}}
\centerline{\epsfig{file=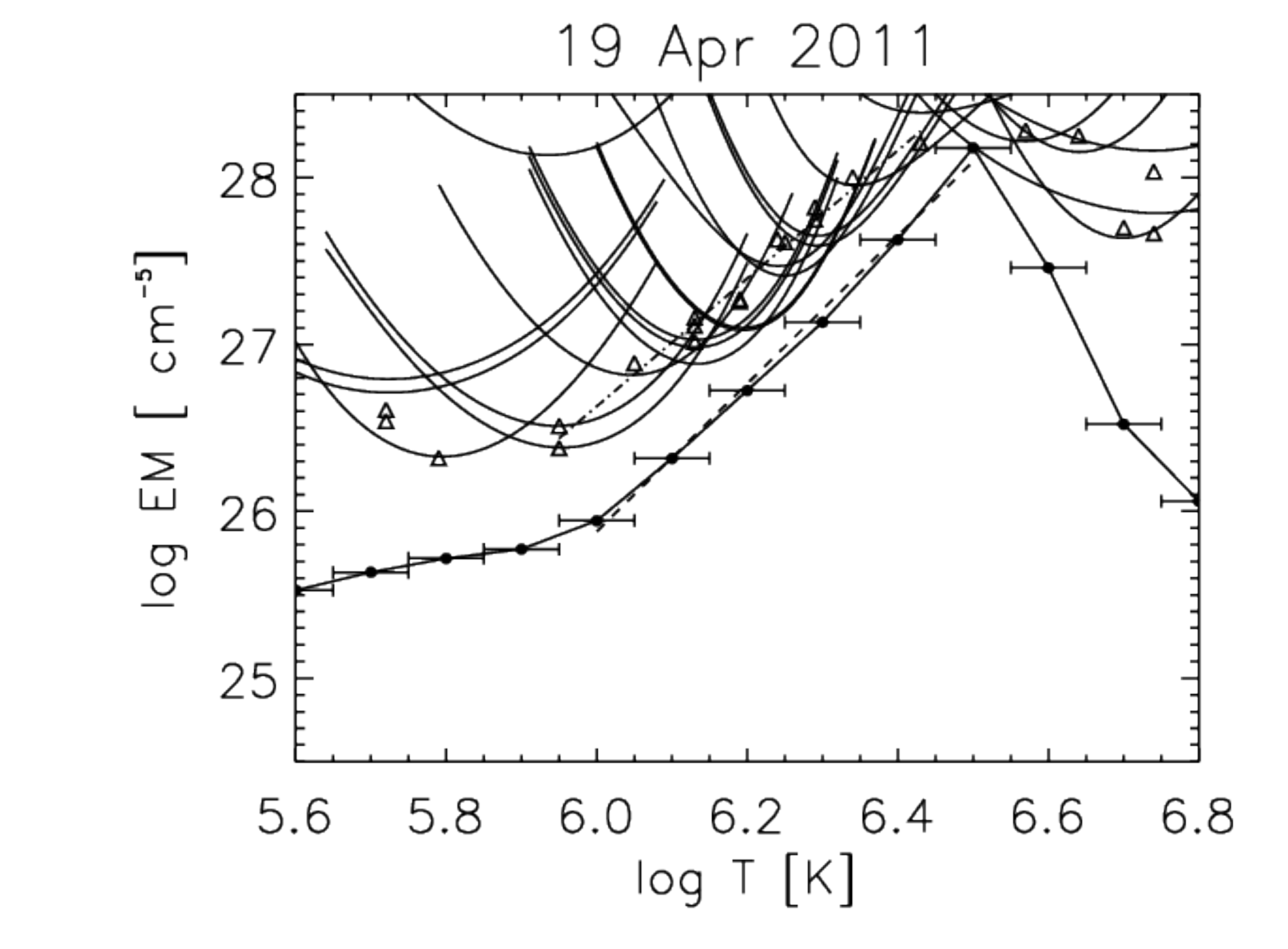, width=3.5cm,angle=90}
\epsfig{file=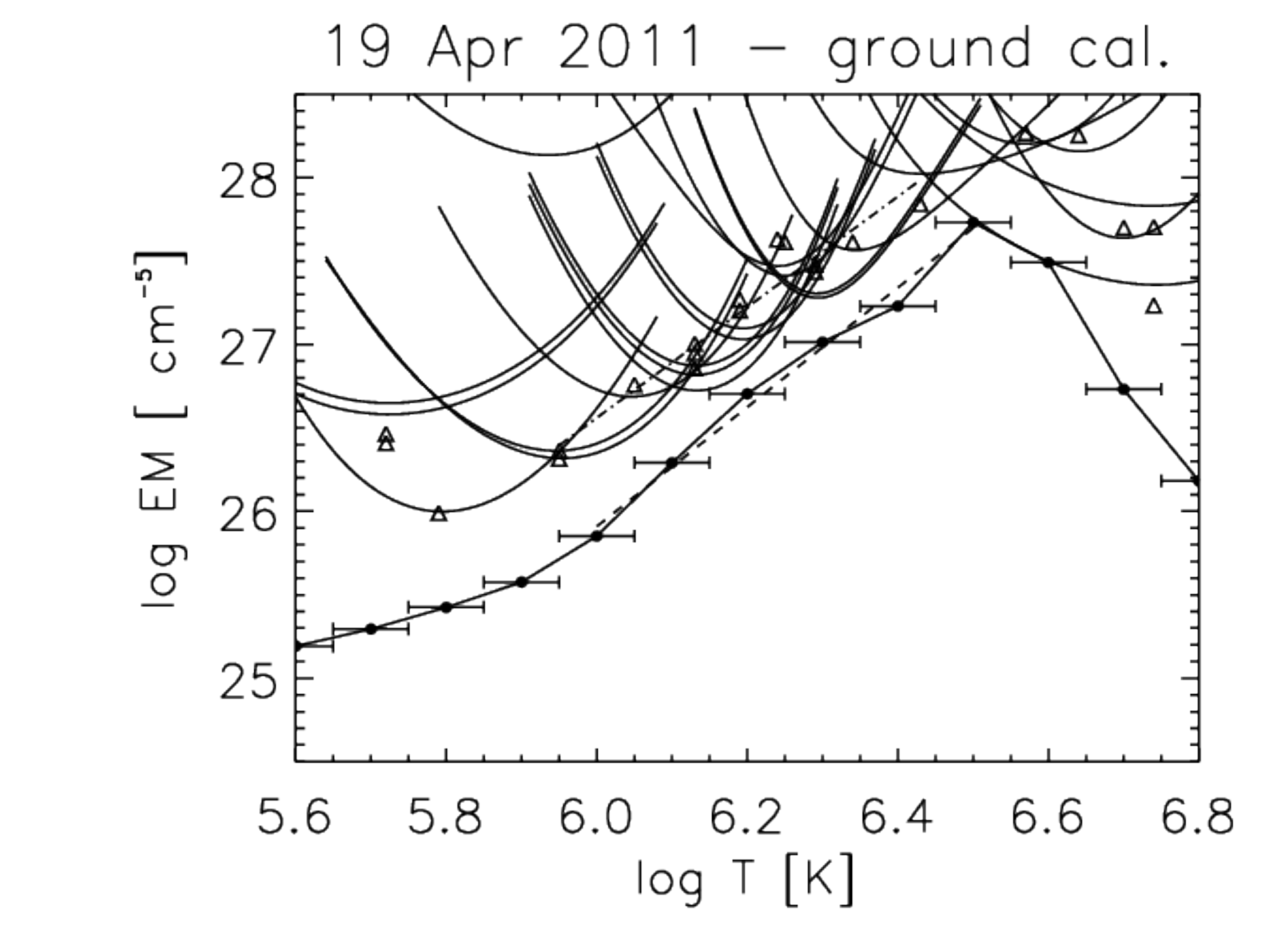, width=3.5cm,angle=90}
}
\caption{From top to bottom: 1) DEM for the 19 Apr AR core, obtained with the spline 
method (black smooth curve), with the MCMC\_DEM program (red curve) and 
XRT\_DEM  method (green curve). 
The points are plotted at their effective temperature, and the values in brackets indicate the ratio
between predicted and observed intensity. 
2)  The results of the Monte Carlo XRT\_DEM inversion.
3) The DEM obtained from MCMC\_DEM.
4) The EM(0.1) values obtained from the DEM, together with the 
curves for the EM loci, and the slopes (dashed lines). 
The  $EM_{\rm jw}$ points (triangles) are also shown, with their 
slopes (dot-dash lines).
The left bottom plot is with the new Hinode EIS calibration (slope=4.4 from the 
 EM(0.1)  and 3.8 from the  $EM_{\rm jw}$ points), the right one
with the ground calibration (slope=3.6 from the 
 EM(0.1)  and 3.3 from the  $EM_{\rm jw}$ points).
}
\label{fig:dem_19apr}
\end{figure}

As shown in  Fig.~\ref{aia_eis_19_apr} (dashed lines), we selected a region within the AR core
for a detailed DEM analysis. We averaged the EIS spectra and obtained 
calibrated radiances.
We measured the electron density using ratios from Fe~XIII (202.0 vs. 203.8~\AA) 
and Fe~XIV (264.7 vs. 270.5~\AA). For both ions, we obtained a  density in the core region of  $4\times10^{9}$~cm$^{-3}$. 
We used this value to calculate the line emissivities for the $DEM$ inversion, although we note that 
the choice of density has little effect on the line emissivities, because of the 
choice of lines.

The results of the three DEM 
inversions methods are shown in the top panel of Fig.~\ref{fig:dem_19apr}. 
We obtain a  distribution with a well-defined peak
at 3~MK, and very good agreement (to within 20--30\%) between observed and predicted intensities.
 There is  good agreement between the three different 
DEM inversions methods in the 1--3~MK region, which is reassuring.

The MCMC\_DEM program  was run with a  temperature grid of log $T$[K]=0.1. 
It tends to underestimate  the peak of the $DEM$ at 3~MK,
 while the emission measure above 3 MK is overestimated. 
If  a finer grid is chosen, the DEM peak tends to agree with the 
spline method, but the DEM in the  1--3 MK range shows large deviations
(for a discussion on the  MCMC\_DEM grid size see  \cite{landi_etal:2012,testa_etal:2012}).
The XRT\_DEM method was run on a finer grid, log $T$[K]=0.05, which produces 
good agreement with the spline method at the peak.

The middle panels of Fig.~\ref{fig:dem_19apr} display the results of the 
Monte Carlo XRT\_DEM and MCMC\_DEM inversions.
The XRT\_DEM  simulations are obtained by randomly varying (400 times) the 
input intensities within the estimated uncertainties, which have been taken as 
20\%, the overall uncertainty in the EIS calibration \citep{delzanna:13_eis_calib},
added to the uncertainty from the fitting. 
The `error bars' on the MCMC\_DEM plots are obtained using the default values
originating from the iterative runs of the program. 
The  Monte Carlo simulations suggest an uncertainty of about 0.4 dex on the EM slopes.

The bottom (left) panel of Fig.~\ref{fig:dem_19apr} displays the $EM(0.1)$ values obtained from
 the spline DEM, 
together with the curves for the EM loci.
 The dashed line indicates a slope of 4.4 for the EM between 1--3~MK. 
We note that if the $EM_{\rm jw}$ points 
are considered, the  slope becomes 3.8,  in agreement with what obtained from just the two 
Fe IX 188.5~\AA\ and the Fe XVI 263.0~\AA\ lines, 
shown in  Fig.~\ref{aia_eis_19_apr}). 
In other words, the EM approximation tends to under-estimate the slope. 
This can be understood by the fact that the \citet{jordan_wilson:1971} 
approximation is close to the loci of the EM loci curves, which are
upper limits to the $EM(0.1)$ values.

We note that \cite{delzanna:13_eis_calib} has shown that the intensities of all the strong EIS 
lines formed in the 2--3~MK range (in the long-wavelength 
channel) have been underestimated by about a factor of two, for data taken after 2009, 
hence most previous analyses have underestimated the slope of the EM in the 1--3~MK range.
 This is because all the strong EIS lines formed around 1~MK are in the EIS short-wavelength channel, which has not 
been affected so much by degradation.
 Fig.~\ref{fig:dem_19apr} (bottom right) shows the EM results obtained form the same data,
 but adopting the ground calibration.
The $EM_{\rm jw}$ slope is 3.3 (ground calibration) instead of 3.8, while the slope
obtained from the $EM(0.1)$ values is  3.6  instead of 4.4.

\subsection{DEM of the AR core for the second rotation}

\begin{figure}[!htbp]
\centerline{\epsfig{file=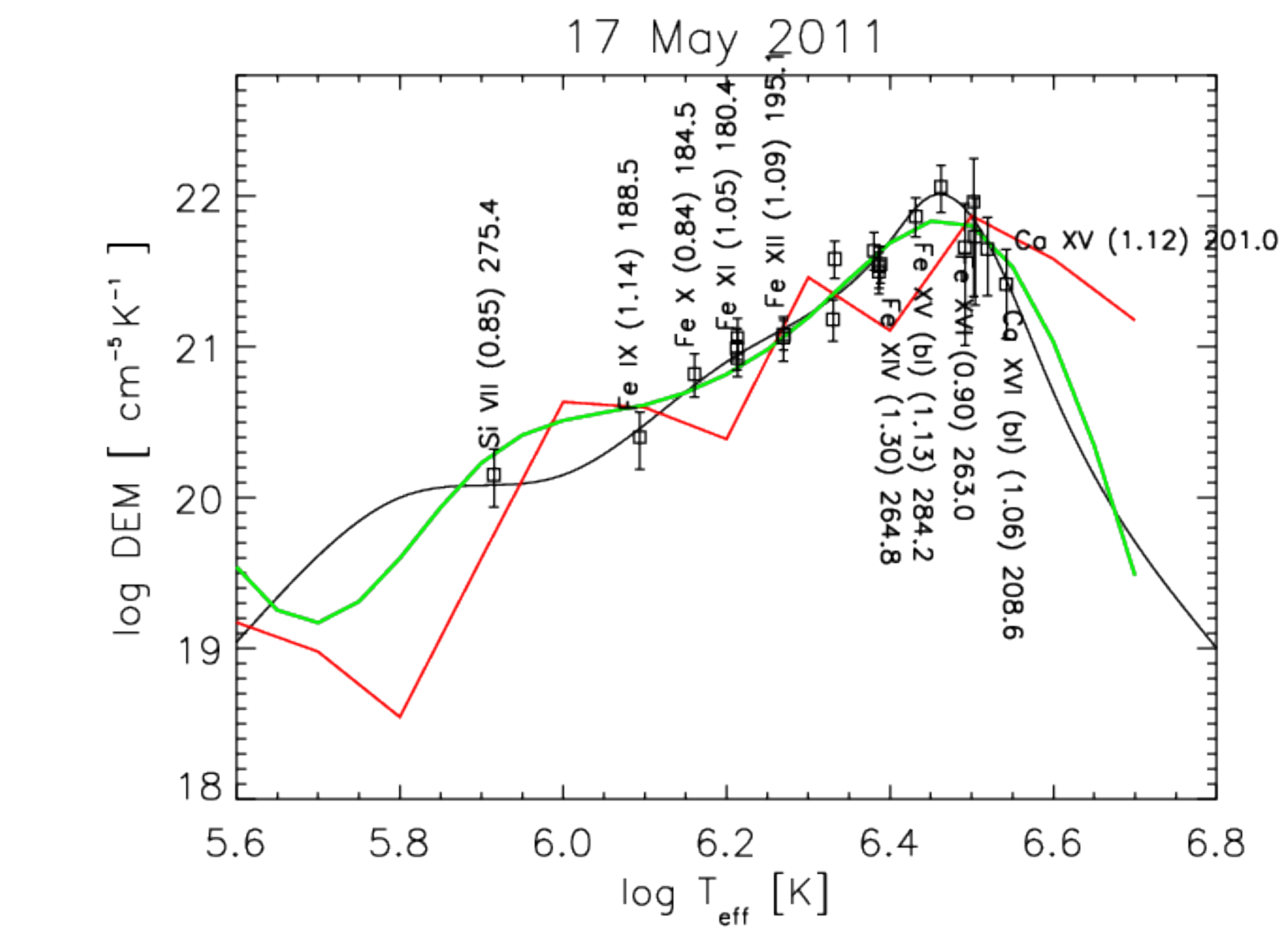, width=6.5cm,angle=90}}
\centerline{\epsfig{file=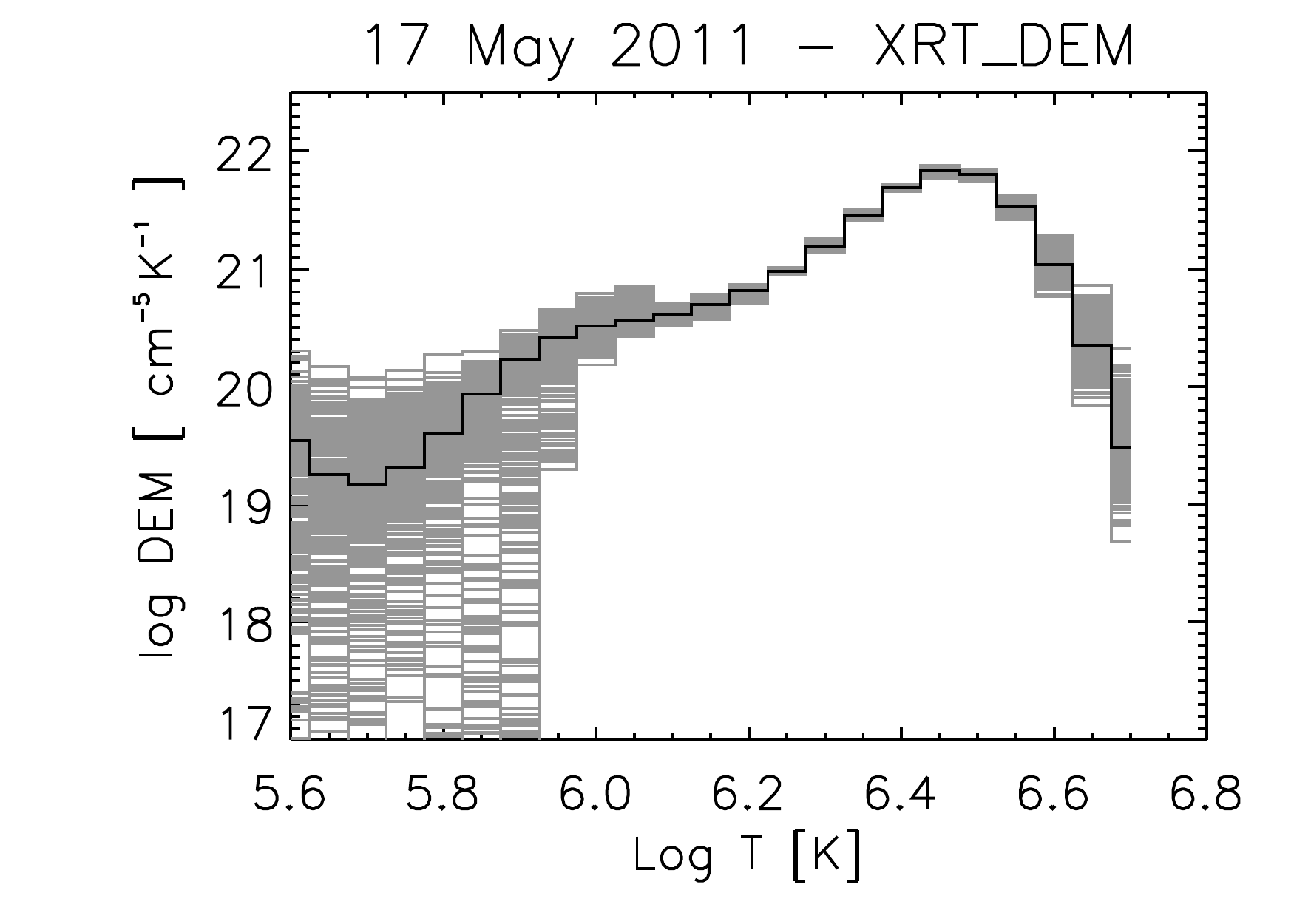, width=6.cm,angle=0}}
\centerline{\epsfig{file=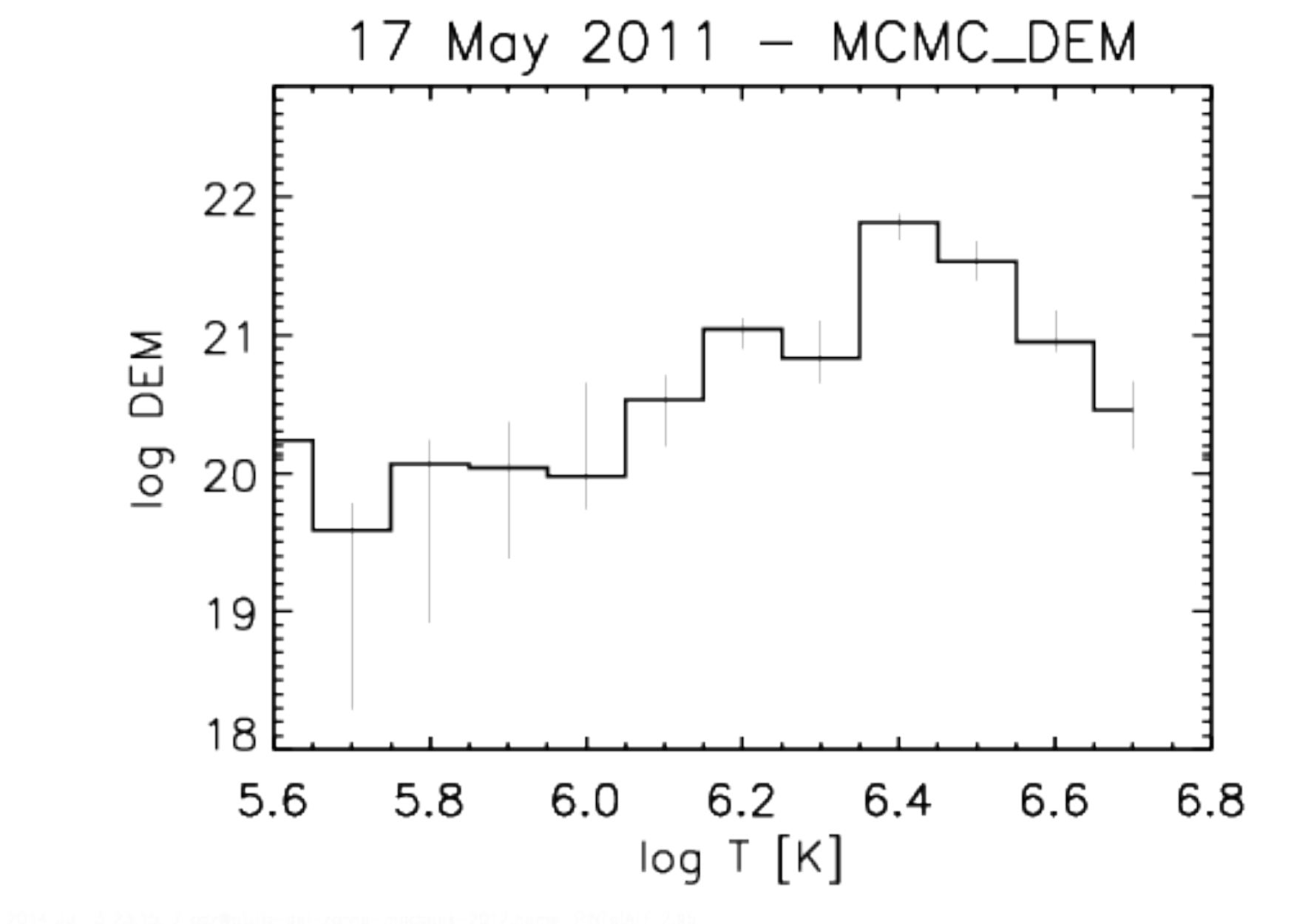, width=6.cm,angle=0}}
\centerline{\epsfig{file=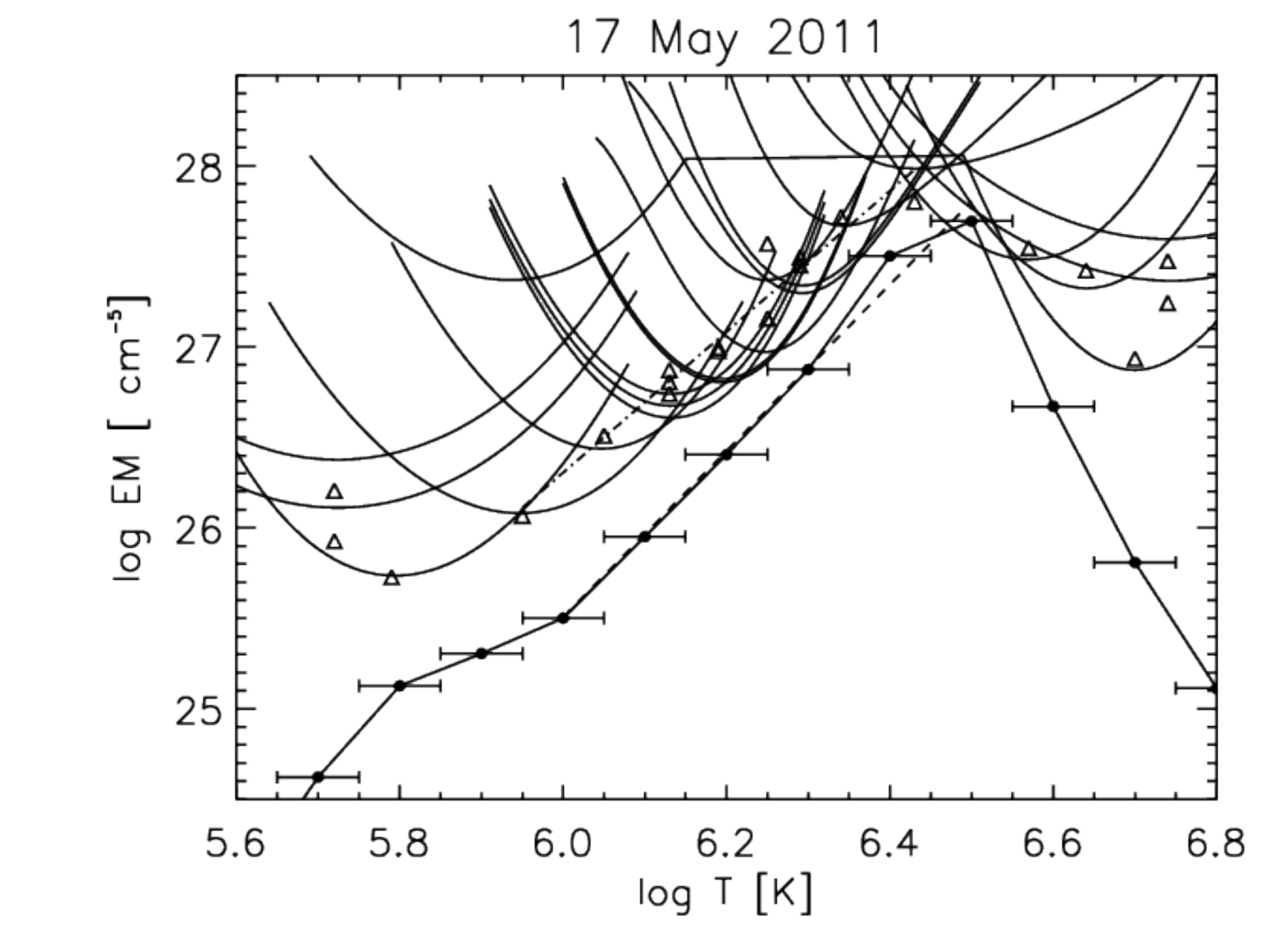, width=4.cm,angle=90}}
\caption{Same as Fig.~\ref{fig:dem_19apr}, for the 17 May 2011 second rotation.
The dashed line in the bottom plot indicates a slope of 4.6 from the EM(0.1), while the dot-dash line
a slope of 3.9 from the $EM_{\rm jw}$ points.
}
\label{eis_dem_17_may}
\end{figure}

As shown in Fig.~\ref{aia_eis_17_may}
(dashed lines), we selected a region within the AR core
for a detailed DEM analysis.
We measured the electron density using ratios from Fe~XIII (202.0 vs. 203.8~\AA) and Fe~XIV (264.7 vs. 270.5~\AA). 
For both ions, we obtained a  
density in the core region of  $2\times10^{9}$~cm$^{-3}$. The density has decreased by a factor of 2 in 
comparison to the first rotation. We used this value to calculate the line emissivities for the $DEM$ inversion.

The DEM and EM results are shown in Fig.~\ref{eis_dem_17_may}. 
Again, there is good agreement among the inversion methods.
The slope from the EM(0.1) values is 4.6, while that from the $EM_{\rm jw}$ points
is lower, 3.9, as we have seen previously. 
We recall that any background subtraction would 
significantly lower the 1~MK  emission, increasing the EM slope.

\section{Summary and Conclusions} \label{conc}

In the present paper, we have studied the EM distribution in the core of active region NOAA~11193, once when it appeared for the first time 
near the central meridian on the visible solar disk 
(when it was 9 days old)
and again when it appeared near the central meridian after a solar rotation. Our analysis 
shows that there is significantly higher emission measure at high temperature in the active region core when the active region is younger. The 
density decreases by a factor of two between the two rotations.

The slope of EM distribution between 1--3~MK, measured using various methods ranged between 2 and 5 throughout the core 
of the active region during the first rotation, with values around 4--5 in the hottest regions. 
In a carefully chosen region inside the core by avoiding any possible contamination from bright moss 
regions  we find a slope of the EM curve to be  4.4$\pm$0.4.

During the second rotation of the active region, the slope between 1--3~MK, measured using  various methods also ranged between 
2 to 5, although in most of the core regions lower values are present.
We find that despite the overall reduction in the EM at 
high temperatures, the slope of the EM between 1 and 3 MK in the hottest regions 
(where Fe XVI is strongest) appears to change little. 
In fact, a carefully selected region without contamination 
from bright moss shows a slope of 4.6$\pm$0.4, i.e. similar to that of the previous rotation.

Additionally, we have also studied  which factors affect the measurement of the slope
 of the EM distribution between 1~MK and 3~MK obtained
from Hinode EIS measurements. The revised EIS calibration has an important effect, by raising the slopes by about 0.8.
The Jordan \& Wilson approximate method  consistently underestimates the slope by about 0.5. 
We have indeed carried out the analysis of one of the regions selected in \cite{TriKM:11} and found a similar 
underestimation. 
 We have tested different inversion methods and have found consistent DEM distributions within  
the 1--3~MK range. 

To the best our knowledge, using AIA and EIS observations for the first time we have shown 
that the slopes significantly vary within 
the cores of an AR,  by about $\pm$1 in the hottest regions. 
There are some slight  differences between the slopes obtained with Hinode/EIS and
SDO/AIA. These  are partly due to the better spatial resolution of
SDO/AIA. Approximate methods tend to underestimate the slopes, compared
to those based on the full DEM inversion.
We therefore suggest that future studies obtain the EM slopes 
after performing a full DEM inversion.

We also pointed out the need to carefully select representative regions, using the much higher 
AIA spatial resolution, to avoid areas where significant  line-of-sight (mostly background) emission is present. 
This is often almost impossible to achieve at the EIS resolution.
During the second rotation, line-of-sight contamination at 1~MK is more
significant than during the first rotation. 
The slopes we provide  are without background subtraction.
Taking the background
subtraction into account would  increase 
the slope of the EM, especially  for the second rotation.
In any case, the slopes in the range 4--5 found for NOAA~11193  
are consistent with high frequency heating occurring 
during the evolution of the active region.

We believe that this study provides an important contribution 
to the problem of coronal heating in active region cores, in
particular by exploring in depth the uncertainties in such analyses.

\begin{acknowledgements}
GDZ, HEM and BOD acknowledge support from STFC. 
BOD also acknowledges support from the Gates Cambridge Trust and Inter-University 
Centre for Astronomy and Astrophysics (IUCAA) for the hospitality during his visit. This work
has benefited from discussions at a series of ISSI workshops. 
 Hinode is a Japanese mission developed and launched by 
ISAS/JAXA, with NAOJ as domestic partner and NASA and STFC (UK) as international partners. It is operated by 
these agencies in co-operation with ESA and NSC (Norway). We acknowledge the use of the SDO/AIA observations for this 
study. The data are provided courtesy of NASA/SDO, LMSAL, and the AIA, EVE, and HMI science teams. CHIANTI is a 
collaborative project involving George Mason University, the University of Michigan (USA) and the University of Cambridge (UK).
We would like to thank Amy Winebarger for useful comments to the manuscript.
\end{acknowledgements}

\bibliographystyle{aa}
\bibliography{references}  

\end{document}